\documentclass[12pt]{article}
\input epsf.sty
\usepackage{epsfig}
\newcommand{\beq}{\begin{eqnarray}}
\newcommand{\eeq}{\end{eqnarray}}

\newcommand{\HH}{{\cal H}}

\newcommand{\OO}{{\cal O}}

\newcommand{\LL}{{\cal L}}

\newcommand{\be}{\begin{equation}}
\newcommand{\ee}{\end{equation}}
\newcommand{\ben}{\begin{eqnarray}\displaystyle}
\newcommand{\een}{\end{eqnarray}}
\newcommand{\refb}[1]{(\ref{#1})}
\newcommand{\p}{\partial}
\newcommand{\pb}{\bar \partial}
\newcommand{\sectiono}[1]{\section{#1}\setcounter{equation}{0}}

\newcommand{\Tr}{{\rm{Tr}}}

\def\sqr#1#2{{\vcenter{\vbox{\hrule height.#2pt
         \hbox{\vrule width.#2pt height#1pt \kern#1pt
            \vrule width.#2pt}
         \hrule height.#2pt}}}}

\begin{document}

{}~ \hfill\vbox{\hbox{hep-th/0312196} \hbox{PUPT-2103} }\break

\vskip 1cm

\begin{center}
\Large{\bf    A paradigm of open/closed duality}

\vspace{0.6cm}

\large{\bf Liouville D-branes and the Kontsevich model }

\vspace{20mm}

\normalsize{Davide Gaiotto and Leonardo Rastelli}

\vspace{10mm}

\normalsize{\em Joseph Henry Laboratories, Princeton University,}

\vspace{0.2cm}

\normalsize{\em Princeton, New Jersey 08544, USA}
\end{center}

\vspace{10mm}

\begin{abstract}

\medskip

We argue that topological matrix models
(matrix models of the Kontsevich type)
are examples of exact open/closed duality.
The duality works at finite $N$ and for
generic 't Hooft couplings.
We consider in detail the paradigm
of the Kontsevich matrix integral
for  two-dimensional topological gravity.
We demonstrate that the  Kontsevich model
arises by topological localization
of cubic open string field theory on $N$ stable branes.
Our analysis is based on standard worldsheet methods in the context
of non-critical bosonic string theory.
The stable branes have  Neumann (FZZT)
boundary conditions in the Liouville direction.
Several generalizations are possible.

\end{abstract}

\newpage

\tableofcontents

\section{Introduction and Summary}

The duality between open and closed
strings is central to modern theoretical
physics.  It underlies, among other things, the
relation between large $N$ gauge theories
and closed strings \cite{thooft}.
Despite impressive progress,
 it is fair to say that we do not yet have a good conceptual grasp
of this correspondence. Even by physics standards,
we are quite far from a ``proof''
of  AdS/CFT  and related examples. We have little
understanding of how general the gauge/gravity
duality is,  let alone how to generate the closed
string dual of a given gauge theory.
With this general motivation in mind, it is  clearly of interest to develop
exactly solvable models of open/closed duality.
An important  class of such models is offered by topological
string theories, the paradigmatic example
being the duality between Chern-Simons
and the closed topological A-model \cite{GV}.

\smallskip

Non-critical strings in low dimensions are another ideal context to
sharpen our understanding of open/closed duality.  Theories
with $c \leq 1$ are fully solvable through the double-scaling
limit of matrix models.\footnote{
For reviews, see \cite{Zinnreview, Igorreview, Gregreview, Joereview}.}
Indeed, the  double-scaled matrix model
for $c=1$ strings has recently been  re-interpreted \cite{JH} as the
``open string field theory''  for an infinite number of D0-branes.
This provides another  beautiful incarnation of
exact open/closed duality.  The doubled-scaled matrix model arises
\cite{Klebanov} as the worldvolume theory of  the  localized Liouville branes.
These are the so-called  ``ZZ branes''  \cite{ZZ},
the unstable Liouville branes
localized in the strong coupling region of the Liouville direction.\footnote{
A similar understanding is available for the double-scaled matrix models of
$c<1$ and $\hat c \leq 1$ theories \cite{chat1, chat2, chat3}.
See \cite{matrix} for recent related work.}

\smallskip

Liouville theory admits also stable branes,
the ``FZZT'' branes \cite{FZZ, T}, which are
 extended in the Liouville direction.
What is the worldvolume theory on such extended branes?

\smallskip

Besides the well-known double-scaled matrix models,
another, more mysterious, class of matrix models
makes its appearance in low-dimensional string theories.
The prototype of these  models, which we shall collectively
refer to as topological matrix models,
is the Kontsevich cubic matrix integral \cite{K},
which computes the exact generating function
of minimal $(2, 2k+1)$ matter coupled to gravity.
Several other examples exist \cite{higherK}, covering
a large class of $c \leq 1$ string theories.\footnote{
The Penner model \cite{Penner}, the $W_\infty$ model \cite{Winf}
and the normal matrix model \cite{normal}
are particularly  intriguing
examples, related to $c=1$ at the self-dual
radius (see \cite{mukhi} for a recent review).}
These models deserve to be called topological because
they  compute certain topological invariants associated
with the moduli space of Riemann surfaces
\cite{WittenTG, K, DW, WittenN,WittenN2, mukhi}.
However, it must be noted that they
actually contain all the information of the
physical theories which are reached from the ``topological point''
by turning on deformations.
As a result, any $(p,q)$ bosonic string theory
admits a polynomial matrix model {\it \`a la} Kontsevich
which completely encodes its exact solution.
Topological matrix models are treated in the usual
't Hooft expansion, with no double-scaling limit.

\smallskip

The reader will have guessed our punchline.  Our basic
contention is that topological matrix models generically
arises in topological non-critical string theories
as the open string field theory on $N$ extended (FZZT)
Liouville branes  (tensored with an appropriate matter boundary state
depending on the string theory under consideration).
 In this paper we work out in detail the prototype of the Kontsevich
model. It is easy to envision
 that several generalizations should exist.
We are going to argue that topological matrix models
are examples of exact open/closed
duality in very much the same spirit as the AdS/CFT correspondence.

\smallskip

Perhaps the most interesting general lesson is that in this exactly
solvable context we will able to precisely describe the mechanism by which a Riemann
surface with boundaries is turned
into a closed Riemann surface.  Open string
field theory \cite{OSFT} on an infinite number of D-branes
is seen to play a crucial role.  Essentially the same
mechanism is at work in the large $N$ transition
for the topological A-model \cite{WittenCS, GV, OV}. The Kontsevich integral
offers an even more tractable case-study.

\smallskip

Very recently, an interesting
 paper has appeared on the archive \cite{Aganagic:2003qj}.
Building on previous work ({\it e.g.}\cite{previous}),
these authors interpret  topological
matrix models as describing the dynamics of non-compact
branes in the topological B-model  for non-compact Calabi-Yau spaces.
Although the language of \cite{Aganagic:2003qj}
is very different from ours, there are clearly
deep correspondences as well.  Understanding in detail the relation between their point
of view and ours should be an illuminating enterprise.

\smallskip

Since the subject of topological matrix model
  may not be very widely known,
and our explicit analysis  will involve  a few
technicalities,  in the rest of this introduction
we review  some background material
 and summarize our main conceptual points.

\subsection{From open to closed worldsheets}

It may be useful to begin  by recalling
the classic analysis \cite{thooft}
of the large $N$ limit of a gauge theory.
In 't Hooft's double line
notation, each gluon propagator becomes
a strip, and gauge theory Feynman diagrams take the
aspect of ``fatgraphs'', or open string Riemann surfaces,
classified by the genus $g$ and the number $h$ of holes (boundaries).
 The generating functional for connected
vacuum diagrams has then the familiar
expansion  (assuming all fields are in the adjoint),
\be \label{thooft}
\log {\cal Z}^{open} (g_{YM}, t) = \sum_{g=0}^\infty \, \sum_{h=2}^\infty \,
(g_{YM}^2)^{2 g -2} \,   t^{h} \, F_{g, h}  \;  ,  \quad t \equiv g_{YM}^2 N \,.
\ee
Nowadays we
interpret this quite literally as the perturbative expansion
of an open string theory, either because
the full open string theory is just
equal to the gauge theory  (as {\it e.g.} for Chern-Simons theory
\cite{WittenCS}),
or because we take an appropriate low-energy limit (as {\it e.g.}
 for  ${\cal N} =4$ SYM \cite{juan}).

\smallskip

The general speculation \cite{thooft}
is that upon summing over the number of holes,
\refb{thooft} can be recast as the genus expansion for
some  {\it closed } string theory of coupling $g_s = g_{YM}^2$.
This speculation is sometimes justified by appealing to the intuition
that diagrams with a larger and larger number of holes look more and more like  smooth closed Riemann surfaces. This intuition is perfectly appropriate for
the double-scaled matrix models,
 where the finite $N$ theory is interpreted
as a discretization of the closed Riemann surface;
to recover the continuum limit, one must send $N \to \infty$ and tune
$t$ to the critical point $t_c$ where diagrams with a diverging
 number of holes dominate.

\smallskip

However, in AdS/CFT,
or in the Gopakumar-Vafa duality \cite{GV},
 $t$ is a free parameter, corresponding on the closed
string theory side to a geometric modulus. The intuition
described above clearly goes wrong here.
A much more fitting way in which the open/closed
duality may come about  in these cases is
for  {\it each} fatgraph of genus $g$ and with $h$ {\it holes}
to be replaced by a closed Riemann surface of  the same genus $g$
and with $h$ {\it punctures}:
each hole is filled and replaced
by a single closed string insertion.
Very schematically, we may  write
\be \label{BW}
 t  \, \int \,    d \rho  \;   \;
 \rho^{\, L_0} \,|{\cal B} \rangle_ P  \leftrightarrow  t \, {\cal W} (P)\, .
\ee
Here the symbol
  $| B \rangle_{P}$ denotes the boundary state creating
a hole of  unit radius  centered around the point $P$ on the
Riemann surface.   Upon integration
over  the length of the boundary (indicated
here by the modulus $\rho$), we can replace the boundary state with a closed
string insertion ${\cal W}$ located at $P$.
 This idea is based on a
correspondence between the moduli space
of open surfaces and the moduli space of closed punctured surfaces
which can be made very precise
(see section 2 of \cite{K}).

\smallskip

Clearly the position $P$  in \refb{BW} is a modulus to be integrated over.
Moreover, summing over the number of holes is  equivalent to exponentiating
the closed string insertion. As a result,
we obtain the operator $\sim \exp (  t \, \int d^2 z \, {\cal W}(z))$,
which implements  a finite deformation
of the closed string background. This is precisely
what is required for the interpretation of $t$ as a geometric
parameter.

\smallskip

We were led to this viewpoint about open/closed duality,
 which probably has a long history (see {\it e.g.}
\cite{Strominger, Green, GV, OV, Shatashvili, GRSZ, GIR, Gopakumar, BOV}),
by thinking about D-branes in imaginary time \cite{GIR}, where
 the mechanism \refb{BW} of boundaries shrinking
to punctures can be described exactly.\footnote{
A closely related viewpoint has been explained
very clearly by Ooguri and Vafa \cite{OV},
using a linear sigma-model perspective.}
In this paper we argue that topological matrix models
are another very precise realization of this idea.

\subsection{Review of {${\bf (2, 2k+1)}$} strings  and  the Kontsevich model}

Minimal bosonic string theories  are specified by a pair $(p,q)$ of
relatively prime integers.\footnote{See \cite{Zinnreview, Dreview, DVVreview}
for reviews and \cite{Nati} for very recent progress
in this subject.}
In the continuum, they are formulated in the usual way by taking
the total CFT = CFT$_{(p,q)} \oplus {\rm CFT}_{Liouville} \oplus {\rm CFT}_{ghost}$.
Here CFT$_{(p,q)}$ is a   minimal $(p,q)$ model  \cite{BPZ},
of central charge
\be
c_{p,q} = 1 -  6 \, \frac{(p -q)^2}{pq} \, .
\ee
The central charge of the Liouville CFT is of course chosen to be
$26-c_{p,q}$ to cancel the anomaly.

\smallskip

The $(2, 2k+1)$ theories will be the focus
of this paper. Perhaps the most familiar among these models is  $(2,3)$, which is pure two-dimensional quantum gravity  ($c=0$),
or string theory embedded in one dimension.
One way to find their
complete solution is by the double-scaling limit of the one-matrix model,
with the potential tuned to the multicritical point of order $k+2$ \cite{DS}.
Each of these theories has an infinite discrete set of physical
closed string   states, conventionally labeled as $\{ \OO_{2m+1} \}$, $m=0,1,2,\cdots$. Observables are correlators of these operators, which is
convenient to  assemble in the following partition function,
summed over all genera $g$,
\be \label{Zdef}
\log {\cal Z}^{closed} (g_s, { t_n}) =\sum_{g=0}^\infty  g_s^{2g-2}
\langle \exp(\sum_{n \, \rm{odd}} t_n {\cal O}_n)  \rangle_g \, .
\ee
The partition functions for the different
$(2, 2k+1)$ theories are connected to each
other by flows of the KdV hierarchy. This means that
we simply need to expand ${\cal Z}^{closed}(g_s, t_n)$ around
different background values of the sources $t_n$ in order to obtain
the correlators of the different $(2, 2k+1)$ models.
 We choose our conventions so that   $\{ t_n = 0 \, , \forall n\}$
corresponds to the $(2,1)$ theory.
 Then correlators for $(2, 2k+1)$ are found by perturbing
 around  $t_n = \delta_{n,3}- \delta_{n,2k+3}$.

\smallskip

As first conjectured by Witten \cite{WittenTG},
the  $(2,1)$ model is  equivalent to
two-dimensional
{\it topological} gravity  \cite{LPW, MS, VV},
superficially a completely different theory.
Topological gravity is a  topological quantum field theory of
cohomological type. In that context,
the operators $\OO_{2n+1} $ are interpreted as Morita-Mumford-Miller classes,
certain closed forms of degree $2 n$  on the moduli space of closed punctured
Riemann  surfaces; correlators $\langle  \OO_{  k_1 }  \cdots  \OO_{ k_n} \rangle_g$ are intersection numbers,
topological invariants of this moduli space.
An index theorem gives the selection rule
\be \label{selection}
k_1 + \cdots + k_n  = 6g - 6 +3 n
\ee
in order for the correlator to receive a non-zero contribution at genus $g$.

\smallskip

The remarkable equivalence of the $(2,1)$ string theory with
topological gravity was proved by Kontsevich \cite{K},
who found a combinatorial procedure to compute these
intersection numbers. Kontsevich  further recognized
that his result for the partition function \refb{Zdef}  could be efficiently
summarized by  the following matrix integral,\footnote{Of
course, as written, the integral diverges. Analytic continuation
$X \to i X$
makes the integral convergent for $Z$ negative definite.}
\ben \label{K}
 {{\cal Z}^{closed} (g_s, t)} &= & \rho({Z})^{-1}  \, \int  [d X]   \, \exp \left(-\frac{1}{{g_s}} \Tr \left[ \frac{1}{2} {Z}
X^2 + \frac{1}{3} X^3 \right] \right)  \, ,
\\
 \rho({Z}) & \equiv &  \int  [d X]  \, \exp \left( -\frac{1}{2
{g_s}} \Tr  {Z} X^2       \right) \,  .\nonumber
\een
The integration is  over  the $N \times N$ {\it hermitian}  matrix $X$.  The
matrix  ${Z}$ appearing in the quadratic term is
another $N \times N$ hermitian
matrix which encodes the dependence on the sources $t_k$ through the dictionary
\be \label{tZ}
t_k    =    \frac{ g_s}{k} \,  \Tr Z^{-k} =
\frac{g_s}{k} \, \sum_{n=1}^N \frac{1}{ z_n^k}  \,  \quad  (k \; {\rm  odd} )\, ,
\ee
where $\{z_n\}$ are the $N$ eigenvalues of $Z$.

\break

\begin{figure}[htbp]
\centering
\epsfig{file=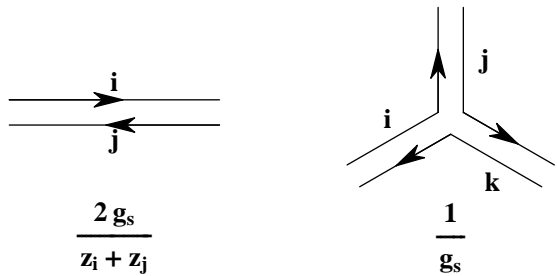, width=3in}
\caption{Feynman rules for the Kontsevich model.
\label{Frules} }
\label{mkplot}
\end{figure}

The Kontsevich integral works in a way which
is truly miraculous - but which  may also strike a familiar chord.
The basic idea is an $n$-point closed string correlator
\be \label{cor}
\langle \OO_{ k_1} \cdots \OO_{k_n}  \rangle_g
\ee
is extracted from the genus $g$ vacuum amplitude
{\it  with $n$ holes}.
 One can proceed perturbatively, using the obvious Feynman rules
that follow from \refb{K} (Figure \ref{Frules}).

\begin{figure}[htbp]
\centering
\epsfig{file=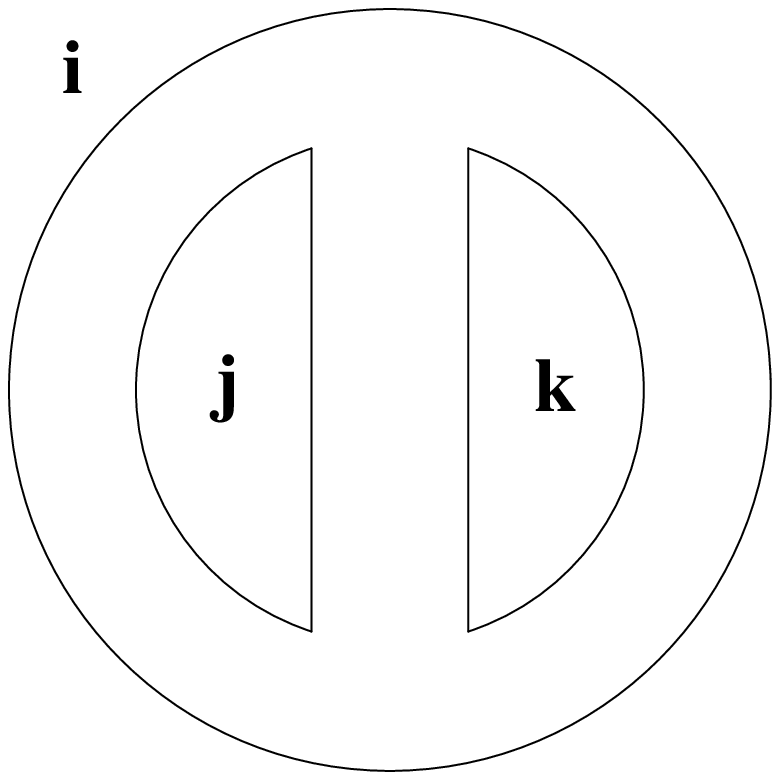, width=1.5in}
\epsfig{file=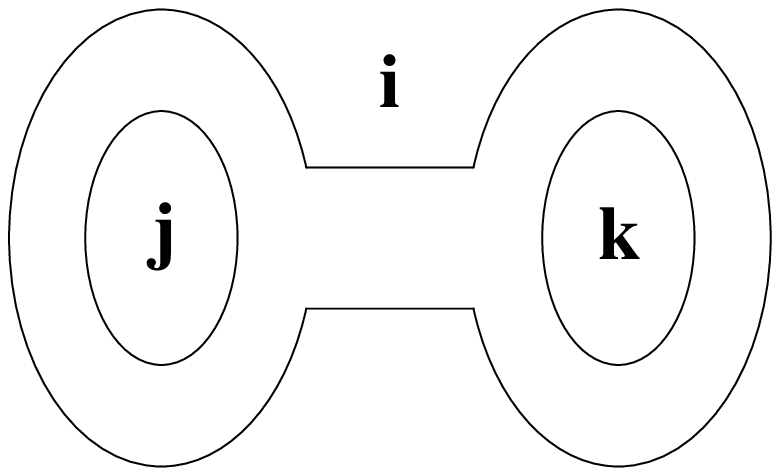, width=2in}
\caption{ The two fatgraphs with $g=0$ and $h=3$.
The indices $i, j, k$ are Chan-Paton labels ranging from 1 to
$N$.  The sum of the two graphs is ${g_s}/(z_i z_j z_k)$.
Upon summing over the Chan-Paton labels, this gives $t_1^3/(6 g_s^2)
\longrightarrow \langle {\cal O}_1 {\cal O }_1 {\cal O}_1 \rangle_{g=0} = 1$.
\label{GammaEx}
}
\end{figure}

Let us define $\Gamma_{g,n, N}$ to be the set
of all connected fatgraphs of genus $g$, $n$ holes,
and a choice of a Chan-Paton index ranging from 1 to  $N$ for each hole
(see examples in Figure \ref{GammaEx}).
The connected vacuum amplitude at genus $g$ and with $n$ holes
 is then
\be    {\cal F}_{g,n, N} =
 g_s^{2g-2+n} \sum_{\gamma \in \Gamma_{g,n,N}}  \frac{1}{\#\, {\rm Aut}(\gamma)}
\prod_{(i,j) \in \gamma}    \frac{2}{z_i + z_j}  \,.
\ee

Individual Feynman diagrams give complicated
rational expressions in the parameters $\{ z_i \}$,  but
remarkably  the total answer  can always be expressed as
\be \label{FC}
  {\cal F}_{g,n, N} = g_s^{2g-2}  \sum^{k_i \; {\rm odd}}_{\{ k_1 \cdots k_n \}}
\frac{
 C_{\{ k_1 \cdots k_n \}}  }
{  \#\, {\rm Aut} ( {   k_1 \cdots k_n   })
}
\prod_{i =1}^n g_s \frac{\Tr Z^{-k_i} }{k_i} \, .
\ee
We see from the definition \refb{tZ} that the parameters
$t_k$ play the role of generalized 't Hooft couplings. From
\refb{Zdef}, we recognize
\be
\langle \OO_{ k_1} \cdots \OO_{k_n}  \rangle_g  = C_{\{ k_1 \cdots k_n \}}   \,.
\ee
The selection rule \refb{selection} is a simple consequence of  Euler's
theorem,
\be
3(2-2g) = 3 (\# V - \# P + n) =- \#P + 3n = -\sum_i k_i + 3n \,,
\ee
where $\# V$ and $\# P$ are the numbers
of vertexes and propagators, and we used that $2 (\# P)= 3 (\# V) $.

\smallskip
Notice that the selection rule \refb{selection} implies that at genus
zero, all two-point correlators vanish. This gives a way to understand
the prefactor $\rho(Z)^{-1}$ in the Kontsevich integral, which
amounts to {\it removing} the fatgraphs with $g=0$, $h=2$
(the annuli) from the vacuum partition function of the matrix model.

\smallskip

In  computing specific correlators using the Kontsevich integral, the rank $N$ can be kept generic,
as long as it is big enough to guarantee that the traces ${\rm Tr} Z^{-k}$
are functionally independent (otherwise the expression
\refb{FC} is not uniquely defined);  $N > {\rm max}(\{k_i\} )/2  $ suffices.
If instead we are interested in the full partition
function ${\cal Z}^{closed}(g_s, {  t_k})$
 for some fixed values of the
infinitely  many sources $ \{ t_k \}$, it is necessary to send $N \to \infty$ in order
for the relation \refb{tZ} to be invertible. So in particular
we need infinite $N$ to compute the correlators of
the higher $(2, 2k+1)$ models, $k >0$.
Nevertheless, it makes perfect sense to keep $N$ finite;
the finite $N$ Kontsevich model covers
an $N$-dimensional submanifold in the moduli space  of the closed string theory.

\subsection{The Kontsevich model is cubic open string field theory }

As we have just reviewed, the correlator  of $n$ closed string operators at genus $g$
is computed in the Kontsevich model by the
fatgraph  vacuum amplitude of genus $g$ and $n$ boundaries.
We propose that this is an exact  open/closed
duality: the Kontsevich model is to be interpreted as
an open string field theory,
dual to the $(2,1)$ bosonic closed string theory.
The Kontsevich integral is to $(2,1)$ string theory
as ${\cal N} = 4$ SYM is\footnote{An apparent
difference is that in AdS/CFT the SYM theory is obtained only in the low-energy limit
of the theory on the D3 branes in flat space,
whereas the Kontsevich model is the full open string field theory. We take
this as a small hint that a better way to understand AdS/CFT should exist, where the SYM theory {\it is}
the  full open string field theory of some appropriate branes. See Section 7.}
 to IIB on $AdS_5 \times S^5$.
 The duality works  just  as explained in section 1.1. The closed string partition function
${\cal Z}^{closed}(g_s, \{t_n\})$
is identified with the vacuum partition function ${\cal Z}^{open}(g_s, \{ z_i \})$
 of the open string field theory.
Each  hole in the open description is
replaced by the insertion of a closed string puncture,
indeed, as we have emphasized in our review of the Kontsevich model,
 powers of the generalized  't Hooft couplings $t_k$ count insertions of the closed string operator ${\cal O}_k$.

\smallskip

The reasoning that led Kontsevich to \refb{K}
uses the decomposition of the moduli
space of Riemann surfaces
\cite{Penner, Harer, Strebel, GMW, BartonProof}
that arises naturally in open string field theory \cite{OSFT} (OSFT),
but  so far this had not been given a direct physical interpretation.
 Here we are saying that in  the Kontsevich model {\it is}  OSFT.
With the advantage of modern insight into the physics
of D-branes, we can give a string theory ``proof''  of Kontsevich result.
The logic is summarized by the following claims:
 \begin{enumerate}
\item One can construct a family of
 stable D-branes in the $(2,1)$ string theory,
labeled by a continuous parameter $z$.
\item Insertion of the boundary state  $| {\cal B} (z) \rangle$ for any one such
brane  in a string amplitude is fully
equivalent to  the insertion of a closed string puncture,  as in \refb{BW}. In this
case, the precise correspondence is
\be \label{BO}
 \int     d \rho \,    \rho^{L_0}  \, | {\cal B} (z) \rangle_P \leftrightarrow
\sum_{k  \, {\rm odd}}  \frac{  {\cal O}_k (P)}{k  \, z^k} \,.
\ee
\item The full cubic OSFT \cite{OSFT}
on a collection of $N$ of these D-branes, reduces
precisely to the Kontsevich action \refb{K}.
The parameters  labeling the branes, $\{ z_i \}$, $i=1 \cdots N$,
are the same as the parameters
appearing in the quadratic term of the matrix integral.
\end{enumerate}
These claims are sufficient to establish Kontsevich result.
We just have to evaluate the string theory
 vacuum amplitude ${\cal Z}$ in the presence
of $N$ branes. We do this in two equivalent ways.
  Evaluating  ${\cal Z}$ in the open channel, we have (claim 3) the sum
of vacuum amplitudes of the Kontsevich integral,  ${\cal Z}^{open}(g_s, \{z_i\})$.
Evaluating ${\cal Z}$ in the closed  channel, we can replace
each hole by a sum of closed string operators (claim 2),
and obtain the generating
function  ${\cal Z}^{closed}(g_s,\{ t_n \})$
 of closed string correlators.
This identifies the vacuum amplitude of the Kontsevich integral
with the closed string partition function,
\be
{\cal Z}^{closed}(g_s, \{ t_n \})  \equiv {\cal Z}^{open}(g_s,\{ z_i \}) \, ,
\ee
which is what Kontsevich showed by more abstract and
rigorous methods.
 The dictionary \refb{BW} between the ``open parameters'' $\{ z_k \}$
and the ``closed parameters''
$\{ t_k \}$  has its microscopic explanation in the rule
 \refb{BO} to replace a boundary with  a specific closed string operator.\footnote{
It makes sense to consider open string vacuum amplitudes
at fixed values  of $\{ z_i \}$ because
 these are superselection parameters that do not fluctuate.
 This statement  is dual to  the  statement
 that the closed string background  $\{ t_k \}$ is superselected
\cite{SS}.}

\subsection{Extended Liouville D-branes in topological string theory}

Our goal is now to justify these claims by standard worldsheet methods.
The $(2,1)$ string theory is strictly speaking outside
the range of  the definition given at the beginning
of section 1.2, since the Kac table is empty and there
is no $(2,1)$ ``minimal'' model. A possible definition is
formal analytic continuation to $k \to 0$ of the double-scaling
results \cite{DS},
but this is unsatisfactory for our purposes.
Fortunately, there are several other more intrinsic formulations,
appearing to all yield the same results.

\smallskip

Since $c_{2,1} = -2$, the simplest choice for the matter CFT is  a pair of free, Grassmann odd scalars $\Theta^1$ and $\Theta^2$.  This provides a continuum definition
of the $(2,1)$ model as $c=-2$ matter coupled to $c=28$ Liouville,
and it is the set-up that we shall use in this paper.
Sitting at the point $\{ t_k = 0 \}$
corresponds in particular to taking the bulk cosmological constant
$\mu \equiv t_1 = 0$.\footnote{It may be useful to recall that
in this theory (unlike the generic $(p,q)$ model, $q \neq 1$)
amplitudes depend analytically on $\mu$ and it
makes sense to treat $\mu$ perturbatively.}

\smallskip

Claim 1 is established by taking Dirichlet
boundary conditions for the $\Theta^\alpha$ and
FZZT boundary conditions in the Liouville direction.
The FZZT boundary state depends on a continuous
parameter $\mu_B$, the boundary cosmological
constant, which can be thought of as the vev of the
open string tachyon living on the brane. We identify $\mu_B = z$.
The full boundary state is then
\be
| {\cal B} (z) \rangle =  |  {\cal B}_\Theta^{\rm Dirichlet} \rangle
\otimes  | {\rm FZZT } (\mu_B = z ) \otimes |{\cal B}_{ghost}\rangle \,.
\ee
FZZT boundary conditions are closely related to the
notion of macroscopic loop operator $w(\ell)$ in
two-dimensional quantum gravity \cite{MSS,AMB}.
 $w(\ell)$ is the operator
that creates a hole of length $\ell$ in the Riemann surface,
where the length is measured with the metric
obtained by taking the Liouville field as the conformal factor.
Then\footnote{Here we are just tensoring
the well-known relation between FZZT branes
and macroscopic loops \cite{MSS, AMB, FZZ} with the
(trivial) Dirichlet b.c. for the $\Theta^\alpha$.}
\be   \label{laplace}
\int    \, d \rho \,  \rho^{L_0} \, |{\cal B }(z) \rangle \sim  \int_0^\infty \, \frac{d \ell}{\ell}\,
 e^{-\ell  \, z}  \, w(\ell) \,.
\ee

\smallskip
To obtain claim 2, we appeal to a standard bit
of lore in non-critical string theory \cite{MSS}.
Under rather general conditions, the  macroscopic loop
operators can be expanded as $\ell \to 0$ as a sum of local
closed string operators,
\be
w(\ell) \sim \sum  \,  \ell^{x_k}\,  {\cal O}_k \, ,
\ee
where $x_k \geq 0$.
A simple argument based on conservation of the Liouville
momentum (section 3.1),
 fixes the exponents to be $x_k = 2k +1$.  The
$\ell \to 0$ expansion of $w(\ell)$ translates after Laplace transform \refb{laplace}
into a $z \to \infty$ expansion of $|{\cal B} (z )\rangle$  as a sum
of terms  $\sim z^{-2k-1} {\cal O}_k  $.
 This gives claim 2,
modulo fixing the precise normalization of the operators ${\cal O}_k$. In
principle these
normalization coefficients could be obtained
 by a very careful analysis of the boundary state, but it it easiest
to determine them indirectly by consistency,
as we explain in section 5.
 This replacement
 of a boundary with a sum of closed string insertions
 is a generic fact
in low-dimensional string theory, and does
not appear to depend on the topological
nature of the $(2,1)$ model.

\smallskip

By contrast, claim 3 is based
on a  mechanism of topological localization,
similar in spirit to the reduction  of  the open topological
A-model on $T^*(S^3)$ to Chern-Simons theory  on $S^3$ \cite{WittenCS}.
The worldsheet boundary CFT
admits a nilpotent scalar supercharge
$Q_S$  \cite{Distler},
anti-commuting with the usual BRST operator $Q_B$.
The open string (first-quantized) Hamiltonian  is a $Q_S$ anti-commutator,
so it can be rescaled  by an overall constant
without changing the physics.  As in the case of
\cite{WittenCS},
the only contributions to open string amplitudes come from
the region of moduli space
where the Riemann surfaces degenerate to ordinary Feynman
graphs. In the usual OSFT decomposition
of moduli space in terms of trivalent vertices
and propagators (strips) of length $t^{(\alpha)}$,
this is the limit in which each $t^{(\alpha)} \to \infty$.
In this limit, the full cubic OSFT collapses to a cubic matrix integral
for the open string ``tachyon''.  A  detailed analysis
of Liouville BCFT correlators
(section 4.3 and appendix) shows that this matrix integral is
exactly the Kontsevich model,
provided we identify  the
boundary cosmological constants $\{ \mu_B^i \}$,  $i=1,
\cdots, N$,  with the parameters $\{ z_i  \}$.

\smallskip

The discussion has been phrased so far in terms of 
worldsheet ideas. An alternative powerful
viewpoint is the use of ``spacetime'' Ward identities,
which we briefly outline in section 5 of the paper.
Finally the whole contruction admits an instructive generalization
to non-zero bulk cosmological constant $\mu$, as
described in section 6.

\sectiono{Closed bosonic strings in ${\bf D=-2}$}

We define the  $(2,1)$ closed string theory
by choosing the total worldsheet action to be
\be \label{S21}
S = S_{matter}^{c=-2} +  S_{Liou}^{c=28} + S^{c=-26}_{ghost} \, .
\ee
The matter CFT is that of a pair of real, {\it Grassmann odd}
 scalar fields $\Theta^1(z, \bar z)$ and $\Theta^2(z, \bar z)$,
with the free action
\be \label{Sth}
 S_{matter}^{c=-2} = \frac{1}{2 \pi} \int d^2 z \, \epsilon_{\alpha \beta}
\partial \Theta^\alpha \pb \Theta^\beta \, ,  \qquad
\alpha,\beta=1,2 \,.
\ee
There is
some freedom as to which CFT with $c=-2$ one
should pick.  Another possibility \cite{Distler} would be to take
 the more familiar  $\xi \eta$ ghost system,
related to the $\Theta^\alpha$ system as follows:
\be
\eta (z) = \p \Theta^2 (z, \bar z)  \, ,
\qquad \xi(z) + \xi(\bar z) = \Theta^1(z, \bar z) \,.
\ee
The two theories differ only in the treatment of the zero-modes.
$\Theta^1(z, \bar z)$  has only
one {\it non-chiral}  zero-mode (the same is true
for $\Theta^2(z, \bar z)$), so the mode expansion reads
\be
\Theta^\alpha(z, \bar z) = \theta^\alpha_0 + \frac{1}{2} \, d_0^\alpha
 \ln|z|^2 +
\frac{1}{\sqrt{2}} \sum_{n=-\infty \, ,
n \neq 0}^\infty \,
\left( \frac{d^\alpha_n}{n z^n}  + \frac{\bar d^\alpha_n}{n \bar z^n}  \right) \,.
\ee
This is a rather subtle difference,
but we believe that the choice of the $\Theta^\alpha$
is the correct one.  First, this is the most
obvious choice to describe ``strings in minus two
dimensions''. It is  indeed the choice singled out by
defining   the theory  from double-scaling
of a matrix model for random surfaces
embedded in minus two dimensions
\cite{minus2, KW1, KW2, KWloops, plefka}.
Second, the treatment of closed string correlators
is simpler, as unlike the $\xi \eta$ system, the $\Theta^\alpha$ system does not require
the introduction of screening charges.
We come back to this point in the next subsection. Finally,
this is the choice that will naturally
lead to the Kontsevich model.

\smallskip

The $\Theta^\alpha$ system has of course
properties very similar to those of a pair of free bosons, one need
only keep track of Grassmann minus signs.  The OPE reads
\be
\Theta^1 (z, \bar z) \Theta^2(0)
\sim -  \frac{1}{2} \log |z|^2 \, ,
\ee
and the stress tensor is
\be
T_{\Theta} =  \epsilon_{\alpha \beta} \p \Theta^\alpha \p \Theta^\beta \,.
\ee
(Note that in this paper we set $\alpha'=1$).  The $\Theta^\alpha$
CFT as an obvious global SL(2) invariance that rotates the fields.
This symmetry does not extend to an affine symmetry but to a $W_3$
algebra \cite{kau}.

\smallskip

It is amusing  to check the modular invariance of the $\Theta^\alpha$ system.
The vacuum amplitude on the torus can be easily
found by explicit computation of the trace,\footnote{To obtain
a non-zero amplitude, we must of course
insert the two zero modes $\theta_0^1$ and
$\theta_0^2$.}
\be
{\rm Tr} \left[ (-1)^F   \, \theta_0^1 \theta_0^2\, q^{L_0 + 1/12}  \, \bar q^{\bar L_0 + 1/12}
\right] =2 \pi  \tau_2 \, |q|^{1/6} \, \prod_{n=1}^\infty
|1-q^n|^4  = 2 \pi \tau_2 \, |\eta(\tau)|^4  \, ,
\ee
and is indeed modular invariant. The unusual factor of $\tau_2$ is a consequence of the
zero-mode insertions, while the $(-1)^F$ factor follows from odd-Grassmanality.
As it should be, this is the inverse of the torus vacuum amplitude
for two free bosons. We should also mention that (orbifolds of)
$\Theta^\alpha$ systems
have been studied in detail \cite{kau}
as prototypes  of  logarithmic CFTs \cite{gurarie, otherlog}.

\smallskip

Liouville CFT  has been  well-understood  in recent years
(see {\it e.g.} \cite{T, FZZ, ZZ} and references therein),
and it is  largely  thanks to this progress that we shall be able to
carry our analysis.  We collect here some
standard formulas:
\ben
&& S_{Liou}   =  \frac{1}{ 2 \pi} \int d^2 z \left(
\p \phi \pb \phi +   Q  \, R \, \phi   + \mu  \,  e^{2 b \phi}\right) \\
&& c_{Liou}  \equiv  1 + 6 Q^2 \, ,  \quad
Q  =  b + \frac{1}{b} \\
&& \phi(z, \bar z) \phi(0)    \sim   -\frac{1}{2} \log|z|^2 \\
&& T^{Liou}  =  -  \p \phi \p \phi + {Q}
\p^2 \phi \\
&& {\cal V}_{\alpha}  \equiv e^{2 \alpha \phi}\, , \quad h_{\alpha} =
\alpha( Q-\alpha) \,. \label{hb}
\een
Specializing to $c_{Liou}=28$, we have  $Q = 3/\sqrt{2}$, $b=1/\sqrt{2}$.
 We shall keep the symbol $b$ in many formulas to facilitate
future generalizations; unless otherwise
stated, it is understood that $b \equiv 1/\sqrt{2}$.

\subsection{Remarks on closed string observables}

In this subsection we offer some side remarks about closed string
amplitudes. Our  main interest is in the {\it open} string sector,
indeed the essential point
is that one can bypass the closed string theory
altogether and compute everything
using open string field theory (the Kontsevich model),
which is structurally much simpler. The subject
of closed string amplitudes in topological
gravity is notoriously subtle \cite{WittenTG, VV, DNcontact, DistlerSemi, DN,
Imbimbo}.
Here we  attempt to make contact with some of the previous work
and suggest that  the action (\ref{S21},\ref{Sth}) may
 offer a different and simpler starting point.

\smallskip

 A  model very similar to (\ref{S21}, \ref{Sth}) (but with the $\xi \eta$
system instead of the $\Theta^\alpha$ system) was considered
by Distler, who observed that by an elegant
change of variables (see   \refb{bos} below)
the bosonic $(2,1)$ theory could be formally
related to the topological gravity formulation of \cite{LPW}.
This is one of the several
\cite{WittenTG,VV,DNcontact,DistlerSemi,DN,Imbimbo}
(closely related) field-theoretic formulations
of topological gravity (see \cite{cordes, weis} for reviews).
They all have in common a sophisticated BRST machinery
extending the ordinary moduli space to a
(non-standard) super-moduli space,
which  in essence is just the space of
differential forms over the bosonic moduli space.
These formulations (as particularly transparent
in Verlinde's set-up \cite{VV}) make it manifest that closed
string amplitudes are intersection
numbers on the moduli space. In this paper we will carry our analysis
in the context of the bosonic $(2,1)$ theory,
but we believe that an analogous
derivation of the Kontsevich model must be possible  in the BRST formulations
of topological gravity.

\smallskip

A potential worry  is  the claim by Distler and Nelson \cite{DN}
that the bosonic $(2,1)$ model ({\it with the $\xi \eta$ system})
does not correctly reproduce the topological
gravity results, and that the full
BRST machinery is necessary to obtain the correct measure of
integration over the moduli space.  It is quite difficult to
compute topological gravity amplitudes from first principles
using standard worldsheet methods,
in any of the field-theoretic formulations.
 The difficulty stems from the very nature of the observables:
amplitudes are naively zero before integration over the moduli space,
and receive contributions only from ``contact terms''
(degenerations of the punctured surface).  This is related
to the fact that there are no non-trivial closed string states
in the absolute BRST cohomology, the only
observables being in the semi-relative cohomology.

\smallskip

However, the different zero-mode structure of the $\Theta^\alpha$
system does certainly affect the calculation
of these contact terms. We believe that a careful
analysis using the action \refb{Sth} would
fully account for the correct contact term algebra.
This is very plausible in light of the fact
that using this worldsheet action we will
obtain the Kontsevich model.  More concretely,
our derivation of the Kontsevich model
suggests a ``canonical''  form for the closed
string vertex operators,
\be
\label{OPk}
 {\cal O}_{2k+1}    =    e^{2(1-k)b \phi}   \, {\cal P} _k (\partial \Theta^\alpha,  \bar \partial
\Theta^\beta)   \, c \bar c   \, .
\ee
Here  ${\cal P} _k (\partial \Theta^\alpha,  \bar \partial
\Theta^\beta)$ is a primary of dimension $\left(\frac{ k(k+1)}{2},
\frac{k(k+1)}{2} \right)$, and it should be invariant under the SL(2) symmetry.
This follows from the fact that the D-branes
which we use to obtain the Kontsevich model are SL(2) invariant.
 It turns out that there is a {\it unique} such operator in the
$\Theta^\alpha$ CFT. This can be seen from the results
in \cite{kau}. In that paper it is proved that
(in each chiral half of the  theory), for each $j \in {\bf N/2}$,
 there is exactly  one  spin-$j$ SL(2) multiplet of primaries,
of conformal dimension $j(2j+1)$. Since there is
only one way to combine the chiral and antichiral
fields into an SL(2) singlet, this shows
the uniqueness of ${\cal P} _k (\partial \Theta^\alpha,  \bar \partial
\Theta^\beta)$.

\smallskip

The operators \refb{OPk} differ from the ones considered
by Distler \cite{Distler}, which are not SL(2) invariant. In \cite{Distler}
a further operation of ``picture changing''  was necessary in
order to obtain non-zero correlators.  In that language,
the operators \refb{OPk} are already in the correct picture and  in principle
their correlators can be evaluated without any extra screening
insertions. The only selection rule comes from anomalous conservation
of  Liouville momentum, and it is precisely \refb{selection}.

\sectiono{Open string theory on stable branes}

We now turn to the open string sector
of the $(2,1)$ theory.
The natural boundary conditions for the $\Theta^\alpha$
system are either Neumann or Dirichlet. Boundary
conditions for the Liouville CFT are either
ZZ (unstable, localized at $\phi \to \infty$)
or FZZT (stable, extended in the Liouville direction).
The choice leading to the Kontsevich model
is to combine Dirichlet b.c. for $\Theta^\alpha$
and FZZT b.c. for Liouville,\footnote{ Another interesting
choice is Neumann for $\Theta^\alpha$
and ZZ for Liouville,  related to the double-scaled
matrix model, see section 6.1.}
\be
\label{FZZTBC} i(\partial \phi - \pb \phi) |_\p = 4 \pi \mu_B
\, e^{b \phi} \, ,    \qquad \Theta^\alpha |_\p  = 0\,.
\ee
The FZZT boundary conditions are  generated by the adding to the Liouville
action the  boundary term
\be \label{Bact}
 \mu_B \int_\partial e^{b \phi} \, .
\ee

\smallskip

One of the basic ingredients of our construction is
the claim that amplitudes with boundaries can
be reduced to amplitudes where each boundary is replaced
by  a specific closed string insertion.
The same phenomenon was demonstrated for
 D-branes in imaginary time \cite{GIR} through a precise
CFT analysis in the usual framework of (critical) string theory.
 In the present case it is easiest to use instead the language
of  two-dimensional quantum gravity (or non-critical string theory).
This language gives a   very useful  geometric understanding of the
FZZT boundary state, which we now review.

\subsection{Macroscopic loops}

In critical string theory, we   are instructed to integrate
the appropriate CFT amplitudes
over the moduli
space of Riemann surfaces.  In quantum gravity, we integrate over the
two-dimensional metric (modulo diffeomorphisms).
Of course the two points of view are completely equivalent,
as the integral over metrics can be replaced
by the Liouville path-integral followed by integration
over the moduli. Schematically,
\be
 \int    \frac{[{\cal D} g_{ab}] }{{\rm Diff}}\,    \int  \, [{\cal D}X] \,
\left(    {\cal O}_1 \cdots {\cal O}_n    \, \right)\,
\leftrightarrow
\int _{{\cal M}_{g, n}}\,  [d \, m]\, \int  \, [{\cal D}X] \,        [{\cal D} \phi] \,   [{\cal D} b] \,   [{\cal D}c] \,
\left(    {\cal O}_1 \cdots {\cal O}_n    \, \right) \,.
\ee
Here ${\cal M}_{g,n}$ denotes the moduli space of closed Riemann surfaces of genus
$g$ and $n$ punctures, $\phi$ the Liouville field, $X$ a collective
label for the matter fields, and $\{ {\cal O}_k \}$  a generic assortment
of local operators. To compute amplitudes in the
presence of $h$ boundaries, in the language
of critical string theory we would  of course integrate
over the moduli space of ${\cal M}_{g,n,h}$ of Riemann surfaces with $h$ holes,
specifying appropriate boundary conditions for all the fields.
  In the language of quantum gravity,
FZZT  boundary conditions have the simple interpretation of
 introducing a  ``weight"  for  each
boundary length $\ell_i$ \cite{MSS, FZZ},
\be
{ \int}    \frac{[{\cal D} g_{ab}] }{{\rm Diff}}\,          e^{- \sum_{i=1}^h \,  \mu_B^i  \, \ell_i[g]} \,     \int  \, [{\cal D}X] \,
\left(     \cdots    \, \right)\,   \equiv
 \langle \,  \prod_i  \left[ \int \frac{d \ell_i}{\ell_i} \, e^{-\mu_B^i \ell_i}  w(\ell_i)  \right] \cdots \rangle\,.
\ee
Here on the r.h.s. we have introduced the definition of the macroscopic loop
operator $w (\ell)$, which is the operator
 creating a boundary of length $\ell$ in the two-dimensional
universe.  Note that we have also left
implicit a choice of boundary
conditions for the matter fields $X$.
Another standard object is the Laplace transform of $w(\ell)$,
\be
W(\mu_B ) \equiv \int \frac{d \ell}{\ell} \, e^{-\mu_B \ell }  w(\ell)  \,.
\ee
In the presence of three or more boundaries, each loop operator $w(\ell)$ can be expanded
in non-negative powers of $\ell$ \cite{MSS},
or equivalently, $W(\mu_B)$ can be expanded in inverse powers of $\mu_B$ \cite{MSS}.
Each term in this expansion represents a local disturbance of the surface,
and is thus equivalent to the insertion of a local operator.

\smallskip

In our case, the expansion  will take the  general form
\be \label{WD}
W^{Dirichlet} (\mu_B) = g_s \, \sum_{ k }^\infty c_k \, \frac{ {\cal O}_k}{ \mu_B^{x_k}}  .
 \ee
The superscript on $W$ is a reminder that we are
imposing Dirichlet boundary conditions for the matter fields $\Theta^\alpha$.
The operators $\{ {\cal O}_{2k+1} \}$ are the matter primaries,
appropriately dressed by the Liouville field,
\be
 {\cal O}_k = e^{2(1-k) b \phi}  {\cal P}_k (\partial \Theta^\alpha, \bar \partial \Theta^\beta) \,.
\ee
To write this expression,  we are using the information that the set of matter
primaries $\{ {\cal P}_k (\partial \Theta^\alpha, \bar \partial
\Theta^\beta) \}$
 of the $\Theta^\alpha$ system have dimensions $\left(\frac{k (k+1)}{2}   , \frac{k (k+1)}{2}   \right)$.
Their explicit expressions can be found in \cite{kau}.\footnote{ There
is in fact a whole   SL(2) multiplet
 of primaries of dimension
$\frac{k (k+1)}{2}$
in each chiral half of the theory. However
 the $\Theta^\alpha$ boundary state is an SL(2) singlet (see \refb{DB}),
and this fixes uniquely $ {\cal P}_k (\partial \Theta^\alpha, \bar \partial \Theta^\beta)$ for
each $k$, as remarked in section 2.1.}
The value of the Liouville dressing follows  as usual by requiring that
the total dimension be $(1,1)$. 

\smallskip

Recall also that we are taking the bulk cosmological
constant $\mu =0$. (For $\mu \neq 0$, dimensional
analysis would dictate the coefficients
$c_k$ to be replaced by functions $c_k(\mu_B^2/\mu)$.)
It is  immediate
to determine the powers of $\mu_B$ in \refb{WD} by conservation
of the Liouville momentum. One has to recall that each
boundary carries a Liouville momentum $Q/2$, and
that each factor of $\mu_B$ carries momentum $b/2$. This
fixes $x_k = 2k+1$.  The normalization coefficients
$c_k$ could also be computed with some effort,
but we shall ignore this here. Consistency
of the contact term algebra (section 5) will be an easier
route to fix normalizations.

\smallskip

Although this logic seems perfectly satisfactory,
it would be nice to have a derivation of the
same result using the  language of {\it  critical}
string theory, treating the Liouville theory as an ordinary CFT,
in the same spirit as the argument given for branes in
imaginary time \cite{GIR}.
 The FZZT boundary state can be written
as an integral over the
continuum spectrum of Liouville momenta
$\frac{Q}{2} + i P$ of
appropriate Ishibashi states,
\be |{\rm FZZT}(\mu_B) \rangle = \int_0^\infty   \, dP \;  \Psi(\mu_B,P)\,
|\frac{Q}{2} + i P\rangle \,.
\ee
It is conceivable that the analyticity properties
of the theory in the complex $P$ plane
may allow a contour deformation that
would pick  up only the poles corresponding to on-shell states in
$\frac{b_0}{L_0}
\left(  |{\rm FZZT}(\mu_B) \rangle \otimes |{\rm matter} \rangle
\otimes |{\rm ghost} \rangle \right) $.
This should reduce the boundary state to the same sum
of on-shell closed string insertions  expected from the
quantum gravity argument.

\subsection{Boundary CFT}

The next logical step is to determine
the spectrum of open strings living on
these  stable branes.

\smallskip

In the open sector of the $\Theta^\alpha$
system with Dirichlet boundary conditions,
chiral and antichiral oscillators $d_n$ and $\bar d_n$ are identified, and
we find  a single copy of the chiral current  $\p \Theta^1$
(the same for $\p \Theta^2$)
without any zero modes.\footnote{
Had we defined the $(2,1)$ string theory using a $\xi \eta$
system, a zero mode for $\xi$ would
survive on the boundary ($\xi_0 \equiv \bar \xi_0$,
but {\it one} zero mode is still there).
This would spoil our construction.}
It is amusing to check this statement by a modular
transformation of the annulus partition function.  For this
purpose we write the boundary state,
\be \label{DB}
| {\cal B}_\Theta^{\rm Dirichlet} \rangle =  \exp \left( { \sum_{n=1}^\infty  \frac{1}{n}   \, \epsilon_{\alpha \beta}
 d^\alpha_{-n}  \bar d^\beta_{-n}  } \, \right) \theta^1_0 \theta^2_ 0 \, |0 \rangle \,.
\ee
The annulus amplitude can be swiftly evaluated,
\be
\langle   {\cal B}_\Theta^{\rm Dirichlet}|    q^{L_0 + 1/12}  \, \bar q^{\bar L_0 + 1/12}  | {\cal B}_\Theta^{\rm Dirichlet}\rangle  =   2 \pi \tilde t   \,   \eta(\tilde t )^2 \, ,  \quad    q \bar q \equiv e^{- 2 \pi \tilde t} \,.
\ee
Modular transformation gives $\eta(t)^2$, which is
indeed the same result obtained by tracing over
the open string spectrum described above,
\be
 \Tr_{open} \left[ (-1)^F  \, e^{-2 \pi t  (L_0+ 1/12) }  \right]   =  \eta^2 (t)  \, .
\ee

\smallskip

The open string spectrum of the Liouville BCFT for FZZT boundary conditions
is known to have the usual primaries
$\{ e^{\alpha \phi} \}$, of dimension $h_{\alpha} = \alpha (Q-\alpha)$
(note the factor of two difference with respect to the bulk
primaries \refb{hb}).
As usual in
Liouville field theory,
the continuum spectrum $\alpha = Q/2 + i P$ corresponds to delta-function normalizable states, while real
exponents $\alpha \leq Q/2$ correspond to local operators and are
used in the dressing of  the matter primaries.

\smallskip

A crucial observation, due to Distler \cite{Distler},
is that Liouville and $c=-2$
matter can be formally
combined into a  $\beta \gamma$  bosonic ghost system  of conformal dimensions $(2,-1)$,
\be \label{bos}
\beta = \p \Theta^1 e^{ b \phi }\, ,
 \qquad \gamma = \p \Theta^2 e^{-b \phi} \,  .
\ee
(Recall that for $c_{Liou} = 28$ the parameter $b \equiv 1/\sqrt{2}$).
Distler applied this construction to each chiral half of
the closed theory, where the Liouville CFT was taken to
be  a free linear dilaton  $(\mu = 0)$.  The validity
of the bosonization formulas \refb{bos} is then
a simple consequence of the free OPEs.
This commuting $\beta \gamma$ system
has conformal dimensions $(2,-1)$,
the same dimensions of the usual anticommuting $b c$ ghost system.
 This makes the topological
nature of the theory intuitively clear. In any open string vacuum amplitude,
the oscillator parts of the $b c$ and $\beta \gamma$ path-integrals will
exactly cancel each other, and we should expect  the only surviving contributions to arise from
classical configurations.
This expectation will be made more precise below. A basic ingredient is the
scalar supersymmetry, or topological charge,
\be Q_S \label{QS}
\equiv \oint J_S(z)\, , \quad
J_S(z) \equiv b(z) \gamma(z) = \oint b(z) \p \Theta^2(z) e^{-b \phi(z)} \, ,
\ee
which obeys
\be Q_S^2=0  \,.
\ee
The usual BRST operator of the bosonic string theory,
\be
Q_B = \oint  c(z) \left(\, T^{matter}(z) + T^{Liou}(z) + \frac{1}{2} T^{ghost}(z)\, \right)  ,
\ee
turns out to be $Q_S$-exact,
\be \label{QSQB}
 Q_{B} = \{ Q_S,    \oint  \frac{1}{2}  \beta(z) c(z) \partial c(z) \} \,.
\ee

Turning on the bulk Liouville interaction $(\mu \neq 0)$ is expected to preserve
the topological nature of the theory, since the Liouville term
is $Q_S$-closed.  Here we keep $\mu = 0$ and
leave a discussion of the more general case $\mu \neq 0 $
to section 6 of the paper. 

\smallskip

Crucially for our purposes,
an FZZT brane with Dirichlet b.c. for the $\Theta^\alpha$
will preserve the total charge $Q_S + \bar Q_S$.
This is obvious for zero boundary cosmological
constant, and holds also for $\mu_B \neq 0$
since the boundary interaction is killed by $Q_S^{boundary}$.
Here we are defining an operator
$Q_S^{bondary}$ acting on  boundary vertex
operators by integrating
the current $J_S + \bar J_S$ on a semicircle
around the boundary operator.

\smallskip

We devote the rest of this section to the computation
of the cohomology of $Q_S^{boundary}$,
a technical ingredient that we shall need
in our analysis of the open string field theory.
There is a slight complication due to the
fact that for non-zero boundary cosmological constant $\mu_B$,
the BCFT is interacting and the action
of  $Q_S^{boundary}$ is non-trivial.

\smallskip

Let us first consider the case $\mu_B =0$. Then
the action of $Q_S^{boundary}$ is just
the same as for the chiral $Q_S$ operator
\refb{QS}
and the cohomology may be readily evaluated.
The task is simplified by the realization
that the cohomology must lie in the kernel
of $L_0$ and of $J_0$, the zero-mode
of an appropriately defined current $J(z)$.
Consider the current\footnote{
No confusion should arise between
the parameter $b \equiv 1/\sqrt{2}$ and the antighost
field $b(z)$!}
\be
J(z) \equiv J_{Liou}(z) - J_{bc}(z) =  \frac{1}{b} \partial \phi \, + :b(z) c(z):  \,.
\ee
$J_{Liou}$ is an anomalous current that counts
the Liouville momentum in units of $b$, for example $e^{b \phi}$
has $J_0$ charge one.  The linear combination $J(z)$ is non-anomalous
and it is $Q_S$-exact,
\be
J(z) = \{ Q_S,  c(z) \beta(z)  \} \,.
\ee
This implies that the cohomology of $Q_S$ is contained
in the kernel of $J_0$.  Indeed $Q_S$ is invertible outside
this kernel. Similarly,  the total energy momentum
tensor is $Q_S$ exact. Indeed using \refb{QSQB}
\be
T(z)  =  \{  Q_B,   b(z)  \} = \{  Q_S,   G(z) \}  \, , \quad    G(z) \equiv  2 \beta (z) \p c(z) - \p \beta(z) c(z) \,.
\ee
Hence the cohomology of $Q_S$ is in the kernel of $L_0$.
These two facts  readily allow to identify the cohomology of $Q_S$ as the states
\be
e^{n b \phi(0)} c(0) \p c(0) \cdots \p^n c(0) | 0 \rangle \, ,
 \qquad e^{-n b \phi(0)} b(0) \p b(0) \cdots \p^n b(0) | 0 \rangle
 \,.
\ee

When we turn on $\mu_B$ the BCFT becomes interacting and the action of $Q_S^{boundary}$ more complicated.
Luckily the operator $e^{-b \phi(z)}$ that appears in $Q_S$ is a degenerate field of level two
for the Liouville CFT, and its
OPEs truncate to two terms,
\be
[e^{-b \phi  } ]\,  [e^{\alpha \phi}] =  [ e^{(\alpha -b)\phi}]  +  C_{-} [ e^{(\alpha + b)\phi}]  \,
\,.
\ee
Hence we can write
\be
Q_S^{boundary} = Q_S^{(0)} + \mu_B^2 Q_S^{(2)} \,.
\ee
Note that for $\mu_B \neq 0$, $Q_S^{boundary}$ does not have
definite $J_0$ charge, but it is a sum of the original charge zero term  $Q_S^{(0)}$
plus a  deformation of charge two $ Q_S^{(2)}$.
($ Q_S^{(2)}$ has charge two under $J_0$ because
it has ghost number minus one and shifts the Liouville momentum of $+b$).
This is a mild deformation of $Q_S^{(0)}$. Nihilpotency of the total $Q_S^{boundary}$ for any $\mu_B$
implies
\be
(Q_S^{(0)})^2=0 \, \quad  \{Q_S^{(2)}, Q_S^{(0)}  \} = 0\,,   (Q_S^{(2)})^2=0 \,.
\ee
As the $J_0$ charge of $Q_S^{(2)}$ is nonzero, this implies that $Q_S^{(2)} = \{ Q_S^{(0)} , \cdots \} $
and hence it acts trivially on $Q_S^0$ cohomology.

\small
We conclude that the  cohomology of $Q_S^{boundary} = Q_S^{(0)} + \mu_B^2 Q_S^{(2)}$ has the same
dimensionality as the one of $Q_S^{(0)}$:
one operator for each ghost number. We will mainly be interested in the ghost number one operator,
the open string ``tachyon'' $ e^{b \phi(0)}c_1 | 0 \rangle$.
It is immediate to check that this state is in the cohomology for any $\mu_B$.
We can repeat the same reasoning also to the BCFT
with different boundary cosmological
constants $\mu_B^i$ and $\mu_B^j$
at the two endpoints of the open string.
The only states of ghost number one in the cohomology
of $Q_S^{boundary}$ are the open tachyons between
brane $i$ and brane $j$,
\be
  e^{b \phi(0)}c_1 | 0 \rangle_{ij} \,.
\ee

\sectiono{Open string field theory and the Kontsevich model}

It is our prejudice that open string field theory (OSFT) \cite{OSFT}
must play a fundamental role in the understanding of open/closed
duality.  The Kontsevich model provides the prototypical example.
In this section we construct the OSFT on $N$ of the stable
branes of the $(2,1)$ string theory, and show how it
reduces to the Kontsevich matrix integral.

\subsection{Generalities}

The OSFT on $N$ D-branes takes  quite generally
 the familiar form
\be \label{OSFT}
S[ \Psi ] = -\frac{1}{g_s}
 \left(\frac{1}{2}   \sum_{ij} \, \langle  \Psi_{ij}, Q_B  \Psi_{ji} \rangle  + \frac{1}{3} \, \sum_{ijk} \langle
\Psi_{ij}, \Psi_{jk} ,  \Psi_{ki} \rangle \right)\,.
\ee
Let us briefly review the basic ingredients of this action,
referring to \cite{background} for background material.
The string field $| \Psi_{ij} \rangle $, $i,j =1, \cdots N$, is an element of the
open string state space $\HH_{ij}$
between D-brane $i$ and D-brane $j$. This
is the full state-space of the matter + Liouville + ghost BCFT.
In classical OSFT, we restrict $| \Psi_{ij} \rangle$ to have ghost number
one (in the convention that the SL(2,R) vacuum
$|0 \rangle$ has ghost number zero).
In the BCFT language, which is the most natural
for our purposes, one uses the state-operator map
to represent string fields as boundary vertex operators.
The  string field $|\Psi_{ij} \rangle$ can be expanded
as a sum over a complete set of vertex operators,
\be
|\Psi_{ij} \rangle =  \sum_\alpha   \,   c_\alpha  {\cal  V}^\alpha_{ij} (0)   | 0 \rangle \,.
\ee
Here  $ {\cal  V}^\alpha_{ij} (0)$ is a vertex operator
inserted at the origin of the upper half plane, with
boundary conditions for brane $i$ on the negative
real axis, and boundary conditions for brane $j$ on the positive real axis.

\smallskip

The   2-point and 3-point vertices are then defined
in terms of BCFT correlators on the boundary (real axis) of the upper half-plane,
\beq
\langle  A, B \rangle  & \equiv &   \langle  I \circ A(0) \,  B(0) \rangle_{{\rm UHP}} \,  , \quad  I(z) \equiv -\frac{1}{z} \\
\langle A, B , C \rangle & \equiv &  \langle f_1 \circ A(0) \, f_2 \circ B(0)  \,  f_3 \circ C(0) \rangle_{
{\rm UHP}} \,.
\nonumber
\eeq
Here $f \circ A(0)$ denotes the conformal transform of the operator $A(0)$ by the complex map $f$.
 The precise form of the maps $f_i(z)$,  which implement the midpoint
gluing of the three open strings, can be found in many places and  will not be important for us.

\smallskip

We also recall that the string field obeys  the reality condition
\be \label{reality}
| \Psi_{ij} \rangle ^*  = | \Psi_{ji} \rangle \, ,
\ee
where the $*$ involution is defined to be \cite{GZ}
\be
* = {\rm bpz}^{-1}  \circ {\rm hc} = {\rm hc}^{-1} \circ {\rm bpz} \,.
\ee
The  operation `${\rm hc}$' is hermitian conjugation
of the state (it sends  bras  into  a kets, with complex conjugation
of the coefficients).  The operation `${\rm bpz}$' sends
a bra into a ket according to the rule
\be
{\rm bpz} ({\cal V }(0) | 0 \rangle ) = \langle 0 | I \circ {\cal V} (0) \,.
\ee

\smallskip

Definition of the quantum theory requires gauge-fixing.
This is customarily accomplished by imposing Siegel gauge
$ b_0 | \Psi \rangle = 0 $. One must introduce
Fadeev-Popov ghosts for this gauge fixing, and in fact, since
the gauge symmetry is reducible, one needs
ghosts for ghosts, and ghosts for ghosts for ghosts, ad infinitum.
It is a famous miracle \cite{BT} that  the full second-quantized gauge-fixed
action + ghosts can be written in the form
\be \label{siegel}
S_{Siegel} =   -\frac{1}{g_s}
 \left(\frac{1}{2}   \sum_{ij} \, \langle  \Psi_{ij},  c_0 L_0  \Psi_{ji} \rangle  + \frac{1}{3} \, \sum_{ijk} \langle
\Psi_{ij}, \Psi_{jk} ,  \Psi_{ki} \rangle \right)\, ,
\ee
where $|\Psi_{ij} \rangle $ is now a string field of {\it unrestricted}
ghost number, obeying
\be
b_0 | \Psi_{ij} \rangle = 0 \,.
\ee
The propagator
\be
\frac{b_0}{L_0} =  \int_0^\infty \,  b_0  \, d t \, e^{-t  \, L_0}
\ee
has the geometric interpretation of building
worldsheet strips of  canonical width  $\pi$ and length $t$.
The Feynman diagrams are fatgraphs built joining these
flat strips at trivalent vertices (with the curvature
concentrated at the common midpoint of the three open
strings).  This gives the famous decomposition of the moduli space
of open Riemann surfaces \cite{Penner, Harer, Strebel, GMW, BartonProof}
which plays a crucial role in Kontsevich construction as well.

\subsection{Topological localization }

The  general OSFT action \refb{siegel} is a very complicated object.
In the critical bosonic string, explicit calculations
are available for some simple perturbative amplitudes.
Off-shell, non-perturbative calculations in the classical theory have
so far been possible only  using numerical methods (level truncation).
In the present case, a drastic simplification occurs
thanks to a mechanism of topological localization.
A precedent of this phenomenon was discovered by Witten for the
topological open A-model on the cotangent bundle
$T^*(M)$, which reduces to Chern-Simons on the three-dimensional manifold
$M$.

\smallskip

The  localization works in the way
familiar for  topological theories of cohomological
type. The nilpotent supersymmetry $Q^{boundary}_S$
(henceforth simply $Q_S$)
induces a pairing of the states of the theory, such that
in a vacuum amplitudes almost all states cancel
pairwise; only unpaired states (the cohomology of $Q_S$)
give a non-zero contribution.  Let us demonstrate
this in a more formal way. We are going to prove
that $Q_S$ is a symmetry of the gauge-fixed
OSFT action \refb{siegel}; moreover
the action is almost entirely $Q_S$-exact,
except for the terms involving only the open string
tachyons between the $N$ branes.  This
reduces the OSFT action to an $N \times N$ matrix integral.

\smallskip

The topological symmetry is defined as
\be \label{t1}
\delta_S | \Psi \rangle = Q_S |\Psi \rangle \, ,
\ee
and it is an invariance of the gauge-fixed action.
 \be \label{t2}
\delta_S  \, S_{Siegel} = 0 \, .
\ee
The formal properties that ensure     this invariance
are
\beq \label{Qclosed}
&& \langle V_2 |  \,  ( Q_S^{(1)}  +   Q_S^{(2)} ) = 0 \,  \\
&&  \langle V_3 | \, ( Q_S^{(1)}  +   Q_S^{(2)} + Q_S^{(3)} )  = 0 \,.\nonumber
\eeq
Here we are regarding  the 2-point and 3-point vertices as elements of
 $\HH^* \otimes \HH^*$ and $\HH^* \otimes \HH^* \otimes \HH^*$,
i.e.,  as bilinear and trilinear functionals on the state
space $\HH = \oplus_{ij} \HH_{ij}$.
These properties are an immediate consequence
of the fact that $Q_S$ is the zero-mode of a conserved current. They
are easily proved by contour deformations on the 2- and 3-punctured
disks that define the vertices (see {\it e.g.} \cite{RZ}).

\smallskip

We can now apply the general formal arguments
 given in section 5 of \cite{Wmirror} to conclude that
 the path-integral localizes over the
 fixed locus of $Q_S$, that is, over the
 subspace of states in the cohomology of $Q_S$.
 A more lengthy derivation is as follows.
 We can  write
\beq \label{Vcoho}
\langle V_2 |  & = &  \langle V_2 |^{Q_S \, {\rm coho} } +   \langle W_2 | \,  ( Q_S^{(1)}  +   Q_S^{(2)} )  \,,   \\
 \langle V_3 | & = &      \langle V_3 |^{Q_S \, {\rm coho}}   +  \langle W_3 | \, ( Q_S^{(1)}  +   Q_S^{(2)} + Q_S^{(3)} )  \,.
\eeq
Here we have defined a cohomology problem
for $Q_S$ in the spaces $\HH^* \otimes \HH^*$
and $\HH^* \otimes  \HH^*  \otimes \HH^*$
in the natural way. Equ.\refb{Vcoho}  is simply the statement
that since the 2-point and 3-point vertices are $Q_S$ closed
\refb{Qclosed},
they can be  written as a sum of a term
in the $Q_S$ cohomology plus a $Q_S$-exact term.
By K\"unneth formula
the cohomology in the  tensor product space is the tensor
product of the cohomology.   Thus, dropping $Q_S$-exact terms,
we can restrict the whole OSFT action to the
string fields in the cohomology of $Q_S$.

\smallskip

The cohomology of $Q_S$ was computed in section 3.2 and
consists of the states
\be
e^{n b \phi(0)} c(0) \p c(0) \cdots \p^n c(0) | 0 \rangle_{ij} \, ,
 \qquad e^{-n b \phi(0)} b(0) \p b(0) \cdots \p^n b(0) | 0 \rangle_{ij} \,.
\ee
Of these states, only the ones with  $bc$ ghost number $\geq 1$
satisfy the Siegel gauge condition.  Among them,
only the open string ``tachyons''
\be
|T_{ij} \rangle \equiv  e^{b \phi} c_1 | 0 \rangle_{ij} \,
\ee
can give a contribution to the action, since all the other
fields do not saturate the conservation of $bc$ ghost number,
which must add up to three.
This concludes the argument that  the  OSFT action
reduces to the terms containing only the open string tachyons.

\subsection{Liouville BCFT and the matrix model}

 Writing the string field $|\Psi_{ij} \rangle$ as
\be
| \Psi_{ij} \rangle = {X_{ij}} {|T_{ij} \rangle } + \cdots
 \ee
for some coefficient $X_{ij}$, $i, j = 1, \cdots N$,
the OSFT reduces to a matrix model
 for the $N \times N$  matrix $X$. The reality condition \refb{reality} for the string field implies that $X$ is hermitian.
The action for the matrix integral is
\be
S[X] =
-\frac{Volume}{g_s} \left({\frac{1}{2} \,  X_{ji} X_{ij}}\, {\langle T_{ji}
, c_0 L_0 T_{ij} \rangle } + {\frac{1}{3}\,  X_{ij}X_{jk}X_{ki}}\,
{\langle T_{ij} , T_{jk} ,  T_{ki} \rangle} \right)\, .
  \ee
Here we are normalizing the inner products so that
\be
\langle c_1,  c_0 c_1 \rangle = 1 \, ,
\ee
and correspondingly we have extracted a factor
of the (divergent) volume of the brane coming from the
integration over the zero mode of the Liouville
field.\footnote{This overall factor is present
also in all the closed string correlation functions
of the the ${\cal O}_k$ operators,
and it will consistenly cancel out in all
formulas.}
It only remains to evaluate the 2- and 3-point vertices
for the open string tachyons, which define
the coefficients in this matrix action.

\smallskip

The structure of the result can be understood
by a simple reasoning.  It turns out that for the specific values
 of Liouville momenta  that we are interested in, the effect of $\mu_B$ can be treated
perturbatively.  The Liouville  anomaly on the
disk  is $Q=3 b$. A correlator in which the
total Liouville momentum adds to three
(in units of $b$) should then not get any correction from the
presence of a boundary cosmological constant.
Since the open string tachyon has Liouville momentum one,
we expect that the cubic vertex can be
evaluated as a free BCFT correlator,
\be \label{T1}
\langle T_{ij} , T_{jk},  T_{ki} \rangle = 1 \,.
\ee
Notice that  the local coordinates $f_i(z)$ play no role since
these are on-shell primary vertex operators.
On the other hand, in  the kinetic term
we expect to need one insertion of the
 boundary cosmological constant  to saturate the
anomaly. This contribution can come from either side of the
strip, so it is reasonable to guess
\be \langle T_{ij} , c_0 L_0 T_{ij} \rangle \sim \mu_{B}^{(i)} +
\mu_{B}^{(j)} \,.
\ee
With these values for the coefficients the OSFT action would then
become \be S[X] = -\frac{1}{g_s} \left({\frac{1}{2}  X_{ij}
X_{ji}}(\mu_{B}^{(i)} + \mu_{B}^{(j)}) + {\frac{1}{3}
X_{ij}X_{jk}X_{ki}} \right)
\ee
This is the Kontsevich model \refb{K}, after the identification
$\mu^{(i)}_B \equiv z_i$.

\smallskip

One may raise an immediate objection to this reasoning:
the kinetic term should actually be zero, since the open tachyon has conformal
dimension zero and is thus apparently killed by $L_0$. Exactly at $c_{Liou}=28$
there is a loophole in this objection, because the scalar product $\langle T_{ij} , c_0 T_{ji} \rangle$ is divergent.
A more careful analysis is then called for, involving the
full machinery of Liouville BCFT.

\smallskip

To regulate the divergence in the tachyon 2-point function,
 we can go slightly off-shell,  considering the state ${e^{(b + \epsilon) \phi} c_1 | 0 \rangle_{ij}}$.
 As we show in the appendix, the  2-point function in boundary
(FZZT) Liouville theory has a pole as
$\epsilon \to 0$,  precisely  with the expected residue,
\be \label{pole}
 \langle e^{{(b + \epsilon)}\phi} e^{({b + \epsilon}) \phi} \rangle_{1,2}
\sim \frac{\mu_{B}^{(1)} + \mu_{B}^{(2)}}{\epsilon} \,.
\ee
This pole cancels the zero from the action of $L_0$,
\be  L_0 \, e^{{\alpha}\phi} c_1 | 0 \rangle =
(\alpha - b)(\alpha - 2 b)  e^{{\alpha}\phi} c_1 | 0 \rangle  =
  \epsilon \, (-b) \,  e^{{\alpha}\phi} c_1 | 0 \rangle \, ,
\ee
giving the desired result.  The careful computation
of the 3-point function (see the appendix)
is rather uneventful and confirms \refb{T1}.

\smallskip
This resonant behavior of Liouville field theory correlators
is related to the fact that the critical exponent
$\gamma_{str} \equiv 1- 1/b^2$  equals minus one.
 In general, a similar resonant behavior occurs
when $\gamma_{str} $ is a  negative integer \cite{KW1}.
The corresponding values of the central charge $c_{Liou} = 1 + 6(p+1)^2/p$,
with integer $p \geq 2$,  are precisely the ones needed to dress the matter
minimal models $(p,1)$.
These are also the models where the string theory is known to be topological
and a matrix model {\it \` a la} Kontsevich exists.

\subsection{Discussion}

We have seen that only on-shell fields (the open string tachyons)
give non-zero contributions. This can
be given a geometric interpretation:
the whole vacuum amplitude has support
on the region of moduli space where
all propagator lengths  in the fatgraph diverge. The localization
on such singular Riemann surfaces is
again familiar from the Chern-Simons example \cite{WittenCS}.
In the language of \cite{WittenCS}, we can say that there are no
ordinary instantons, and only virtual instantons
at infinity  contribute.
It is well-known that in topological gravity
closed string amplitudes are localized
on singular surfaces \cite{WittenTG, VV}.
Here we are
seeing this phenomenon in the  open channel.
While in the closed channel contact terms are quite intricate,
the open string moduli space is structurally much simpler, and
open string contact terms arise only when boundaries touch each other or
pinch. This geometric intuition could
be used to streamline the combinatorial proofs \cite{WittenOn, itzW3}
of the Virasoro constraints for the Kontsevich model.

\sectiono{Open/closed duality and Ward identities}

The main conclusion to draw is that in this theory,
the effect of D-branes can be completely accounted for
by turning on a simple source term for the closed strings,
\be
{\cal Z}^{open}(g_s, \{ z_i \}) = {\cal Z}^{closed}\left(g_s, \left\{  t_k =   g_s \,   \sum_i   \frac{1}{k \, z_i^k}  \right\} \right) \,.
\ee
This conclusion can be strengthened
 by considering the partition function
of the theory in the presence of both a D-brane
and a non-trivial closed string background.\footnote{After submitting the first
version of this paper, we learnt that a similar
approach as the one outlined in this section
was already developed in the ``old" days of matrix models in the interesting
works \cite{J1, JE} (see also the recent paper \cite{JN}). 
We thank  C. Johnson for pointing out these references to us.}

\smallskip
Recall that in the closed string theory, the partition function
is completely determined by the Virasoro Ward identities \cite{VV, DVV}
\beq
{\frac{\p }{\p t_{1}}} {\cal Z} = {{\cal L}_{-2}} {\cal Z} &  \equiv & \frac{t_{1}^2}{2 g_{s}^2}{\cal Z} +
\sum_{k=0}^{\infty}
(2k+3) t_{2k+3} \frac{\p {\cal Z}}{\p t_{2k+1}}  \nonumber \\
{\frac{\p }{\p t_{3}}} {\cal Z} = {{\cal L}_{0}} {\cal Z} & \equiv & \frac{1}{8}{\cal Z} +
 \sum_{k=0}^{\infty}
(2k+1) t_{2k+1} \frac{\p {\cal Z}}{\p t_{2k+1}} \\
 {\frac{\p }{\p t_{2n+5}}} {\cal Z} ={{\cal L}_{2 n+2}} {\cal Z} & \equiv &  \sum_{k=0}^{\infty}
(2k+1) t_{2k+1} \frac{\p {\cal Z}}{\p t_{2k+2n+1}} + \frac{g_{s}^2}{2}
\sum_{k=0}^{n}\frac{\p^2 {\cal Z}}{\p t_{2k+1} \p t_{2n-2k+1}} \,.\nonumber
\eeq
Each of these equations details how  a specific ${\cal O}_k$
operator, when integrated over the Riemann surface, picks
contributions from collision with other operators
or with nodes  of the surface \cite{VV, DVV}.
The second term in the $\LL_{-2}$ and $\LL_0$ equations, and the first
term in the $\LL_{2n+2}$ equation, represent
the collision of two operators. The last term in the $\LL_{2n+2}$ equation
represents the collision between an operator and a node.
(The first term in the $\LL_{-2}$ equation
accounts for the conformal Killing vectors of the sphere, and
similarly the first term in the $\LL_0$ equation accounts for the
CKV of the torus.) The structure of these equations is strongly constrained
by self-consistency; it is only because
the ${\cal L}_{2n}$ form (half) a Virasoro algebra that these equations
have a solution.

\begin{figure}[htbp]
\centering
\epsfig{file=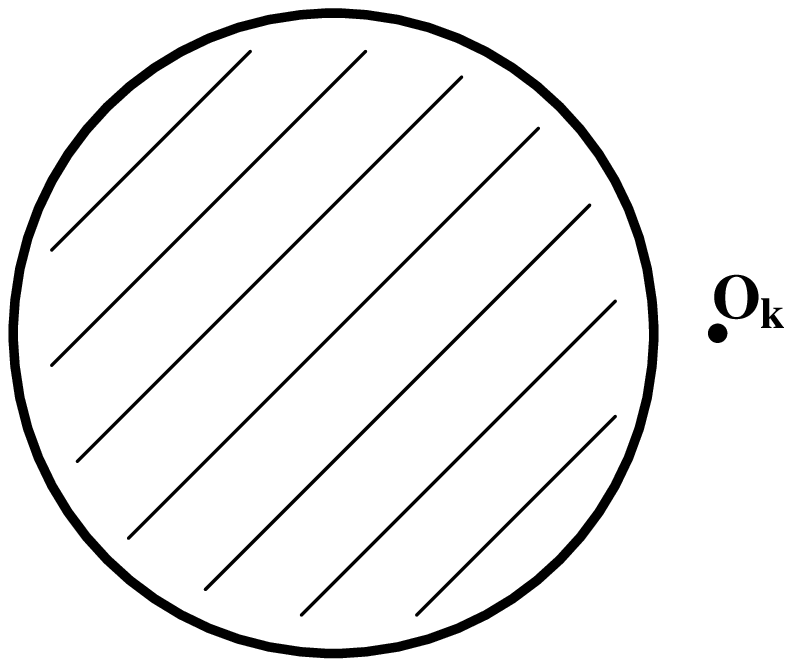,width=1.5in}
\hspace{1cm}
\epsfig{file=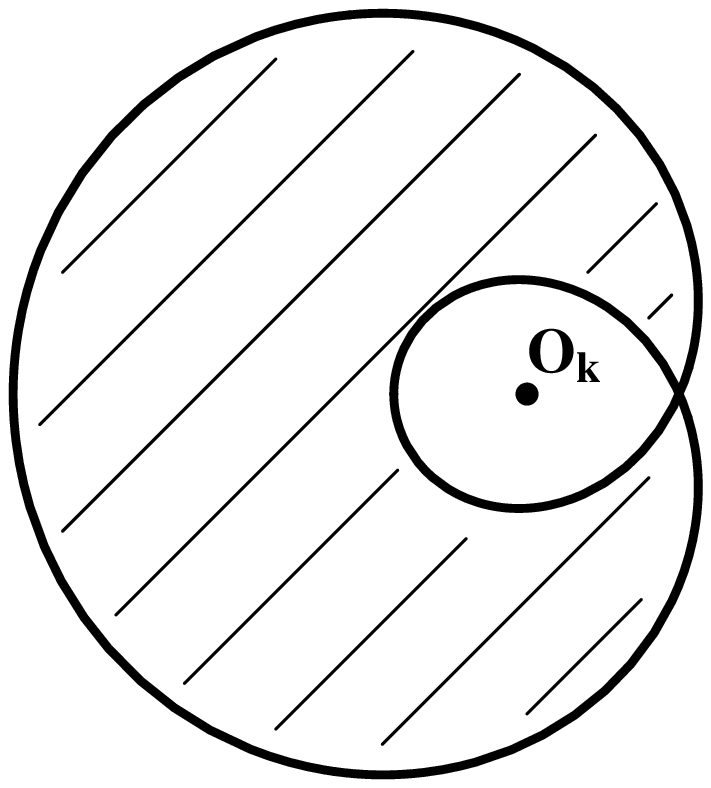,width=1.5in}
\hspace{1.5cm}
\epsfig{file=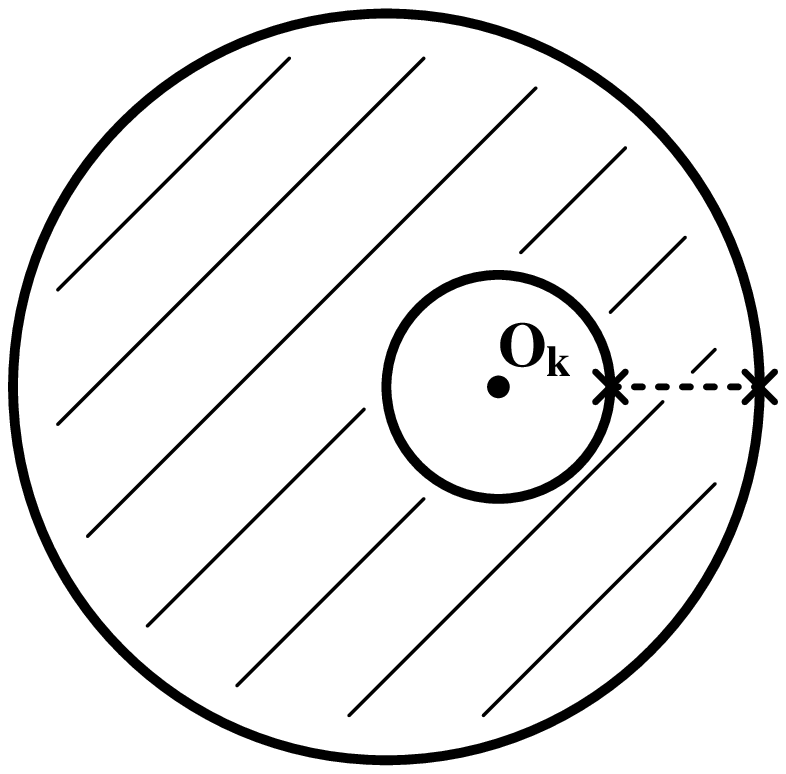,width=1.5in}
\caption{Degeneration of the Riemann surface as the closed string operator
$\OO_k$ approaches the boundary. The shadowed region represents
the hole. As the short neck pinches, the surface
factorizes into two surfaces, each with the extra insertion of an open
string tachyon, indicated by a cross.
\label{collision} }
\end{figure}

\smallskip
To find the partition function
when both D-brane sources and closed string sources
are turned on, we will now extend
these Ward identities by adding the contact terms
that arise from the new ways the surface can degenerate:
when an operator ${\cal O}_k$ collides
with a boundary; and when a boundary collides with a node.
 The collision of an operator
with a boundary
has the schematic aspect shown in Figure
\ref{collision}.

The short neck of the pinching surface
is conformally equivalent to
the insertion of a very long open string propagator;
 the collision leaves behind an open string
tachyon insertion,
with a power of $z$ fixed by conservation of the Liouville momentum.
This piece of knowledge, together with the requirement
that we still have a Virasoro algebra,
uniquely fixes the open + closed  Ward identities.
Considering for simplicity
the case of a single D-brane with parameter $z$,
they have the following form:
\beq{\frac{\p }{\p t_{1}}} {\cal Z} = { \tilde {\cal L}^{(z)}_{-2}} {\cal Z} &\equiv &{{\cal L}_{-2} {\cal Z}} +\left( \frac{t_{1}}{z g_{s}}+\frac{1}{2 z^2} \right) {\cal Z}  - \frac{1}{z}\frac{\p
{\cal Z}}{\p z} \nonumber \\
{\frac{\p }{\p t_{3}}} {\cal Z} = { \tilde {\cal L}^{(z)}_{0}} {\cal Z} & \equiv & {{\cal L}_{0} {\cal Z}} - z \frac{\p {\cal
Z}}{\p z}\\
{\frac{\p }{\p t_{2n+5}}} {\cal Z} = { \tilde {\cal L}^{(z)}_{2 n+2}} {\cal Z} &\equiv &{{\cal L}_{2 n+2} {\cal Z}} -
z^{2n+1}\frac{\p {\cal Z}}{\p z} - g_{s} \sum_{k=0}^{n}
z^{2k+1}\frac{\p {\cal Z}}{\p t_{2n-2k+1}} \,. \nonumber
 \eeq
The terms involving  $\frac{\partial {\cal Z}}{\partial z}$
represent the collision of an operator with a boundary.
The last term in the  ${ \tilde {\cal L}^{(z)}_{2n+2}}$
equation represents the collision of a boundary and a node.
Finally the second term in the   ${ \tilde {\cal L}^{(z)}_{-2}}$
equation accounts for the CKV of the disk with two closed
punctures and of the annulus with one closed puncture.

\smallskip

These identities are sufficient to completely determine
the open + closed partition function ${\cal Z}^{open + closed} (g_s, \{ t_k \} , \{ z_i  \})$.
Not surprisingly, one can easily verify that the
solution is
\be
{\cal Z}^{open + closed} (g_s, \left\{ t_k   \right\} ,\left\{  z_i \right\} )
 =  {\cal Z}^{closed}\left(g_s, \left\{ t_k  + g_s \,   \sum_i   \frac{1}{k \, z_i^k}  \right\}   \right) \,.
\ee
This shows that even when there are non-trivial
closed string sources to begin with,
D-branes can still be re-absorbed into a shift of these sources.
This argument also fixes the overall normalization
in the relation between $t_k$ and $\sum_i z_i^{-k}$.
The closed operators ${\cal O}_k$ have
an intrinsic normalization fixed by the algebra
of closed contact terms. The algebra
of open/closed contact terms can then be used to
fix the coefficients of these canonically normalized ${\cal O}_k$
in the expansion  of the boundary state.
This ties a loose end in our derivation of the Kontsevich model.

\smallskip

We can also define an open partition function in a non-trivial
closed background by subtracting the purely closed
amplitudes,
\be
{\cal Z}^{open} (g_s,  { z_i  }  \, |\,  {  t_k} ) = \frac{{\cal Z}^{open + closed} (g_s, {t_k} , {z_i} )}{{\cal Z}^{closed} (g_s, { t_k} )} \,.
\ee
An interesting question is whether this open partition
function is computed by an appropriate generalization
of the Kontsevich matrix model.
In the next section we provide the answer
for the background with $\mu = t_1 \neq 0$ and $t_k =0$, $k >1$.

\sectiono{Non-zero bulk cosmological constant}

As shown in the appendix, the matrix model obtained by topological
localization of the OSFT action
depends only on the boundary cosmological constants $\{ \mu_B^i\}$
and not on $\mu$.  Clearly however the open + closed string partition function ${\cal
Z}^{open+ closed}(g_s, \left\{ t_k = \mu \delta_{1,k} \right\} | z_i)$ has
a non-trivial $\mu$ dependence. The resolution of this apparent contradiction
is that as we turn on $\mu \neq 0$,
 we must change the dictionary between
the open moduli (the values of the boundary cosmological
constants $\{ \mu_B^i \}$) and the closed
 moduli $\{  t_k\}$. Indeed, the relation
 used so far is based  on the expansion \refb{WD} of the boundary state,
 which - as written - is valid only for $\mu = 0$. 
 
 \smallskip
 As we turn on $\mu \neq 0$,
 we use conventions where
 the parameters $ \left\{ z_i  \right\}$
 are related to the sources $\{ t_k \}$ just as before,
 \be
 t_k = \frac{g_s}{k} \sum_{n=1}^N \frac{1}{z_k^n}  \, , 
 \ee
 but we do not identify anymore $z$ with $\mu_B$,
 rather $z= f(\mu, \mu_B)$ for some function  $f$ which we  now 
 proceed to determine. 
 
 \smallskip
 
To this end, we use the Ward identities derived in
 the previous section. The free energy
\be 
{\cal F}(g_s,\mu, \{ z_i  \} ) \equiv \log({\cal Z}^{open+ closed}(g_s,\{ t_k 
= \mu \delta_{1,k}\} \, |\, \{ z_i \}))
\ee
satisfies
\be \frac{\p }{\p \mu} {\cal F} + \sum_i \frac{1}{z_i}\frac{\p
{\cal F}}{\p z_i} = \frac{\mu^2}{2 g_{s}^2} +\sum_i \frac{\mu}{z_i
g_{s}}+ \sum_{i,j}\frac{1}{2 z_i z_j}  \,.
\ee
This equation can be readily integrated. One finds
\ben \label{Foc}
 {\cal F}(g_s,\mu, z_i) &  =&   \frac{\mu^3}{6 g_s^2}+ \sum_i \frac{1}{g_s}
\left[\frac{1}{3}(z_i^2-2\mu)^{\frac{3}{2}}-\frac{z_i^3}{3} + \mu z_i \right]
+ \frac{1}{2} \sum_{i,j} \log \frac{z_i+z_j}{(z_i^2 -
2\mu)^{\frac{1}{2}}+(z_j^2 - 2\mu)^{\frac{1}{2}}}  \nonumber
\\
&& + {\cal F}(g_s,0, (z_i^2 -
2\mu)^{\frac{1}{2}}) \,.
\een
The first three terms in this expression
represent respectively the changes  of the sphere, disk and annulus
amplitudes as we turn on $\mu$.
The last term  is the sum of all vacuum diagrams
with at least two holes, given as usual by the Kontsevich matrix
integral \refb{K}, but with the replacement $z\to (z^2 - 2 \mu)^{\frac{1}{2}}$.
From the analysis in the appendix we know
that the kinetic term in the Kontsevich integral
is to be identified with the boundary cosmological constant even
for $\mu \neq 0$, hence we learn
$
\mu_B = (z^2 - 2 \mu)^{\frac{1}{2}} $, which gives
the sought relation
\be \label{mod}
 z = (\mu_B^2  + 2 \mu)^{\frac{1}{2}} \,. 
\ee
So far we have argued that
consistency of the theory demands this new relation
between open and closed moduli, which is 
forced upon us by the open/closed integrable structure.
Now we wish to give an independent check of this logic,
and in the process obtain a more physical interpretation.

\smallskip

We start with the Kontsevich representation of the partition function,
\ben  \label{Znew}
{\cal Z}(g_s,0,  (z_i^2 - 2\mu)^{\frac{1}{2}})  & \equiv & \exp(  {\cal F}(g_s, 0,  (z_i^2 - 2\mu)^{\frac{1}{2}}  )  )
\\
& = & 
\rho({(Z^2 - 2 \mu)^{\frac{1}{2}}})^{-1}  \, \int  [d X]   \, \exp
\left(\frac{1}{{g_s}} \Tr \left[ -\frac{1}{2} {(Z^2 - 2
\mu)^{\frac{1}{2}}} X^2 + \frac{1}{6} X^3 \right] \right) \, , \nonumber
\een
and perform the shift $X \to X + (Z^2 - 2 \mu)^{\frac{1}{2}} - Z$ in the
integration variable. This gives
\ben {\cal Z}(g_s, 0 ,  (z_i^2 - 2\mu)^{\frac{1}{2}})  &= &  \exp( - {\cal F}^{D^2})
\cdot \rho({(Z^2 - 2 \mu)^{\frac{1}{2}}})^{-1}  \\
&&  \int  [d X] \, \exp \left(\frac{1}{{g_s}} \Tr
\left[ -\frac{1}{2} {Z} X^2 + \frac{1}{6} X^3 + \mu X \right]
\right) \, ,\nonumber \een
where ${\cal F}^{D^2}\equiv  \left[\frac{1}{3}(z_i^2-2\mu)^{\frac{3}{2}}-\frac{z_i^3}{3} + \mu z_i \right]$
 is exactly the second term in \refb{Foc}.  We observe
 that all the terms conspire to give a simple
 expression for the full open/closed partition function,\footnote{Notice that the third term in \refb{Foc} is precisely taken into 
 account by the $\rho$ prefactors.}
\ben \label{ZX}
{\cal Z}(g_s,\mu, z_i)  &= &  
\exp\left(   \frac{\mu^3}{6 g_s^2}  \right)\cdot\\
&&   \rho(Z)^{-1}  \int  [d X] \, \exp \left(\frac{1}{{g_s}} \Tr
\left[ -\frac{1}{2} {Z} X^2 + \frac{1}{6} X^3 + \mu X \right]
\right) \, .\nonumber
\een
This final equation has a transparent interpretation. Apart
from the purely closed contribution $\exp\left(   \frac{\mu^3}{6 g_s^2}  \right)$
coming from the sphere, the partition function
is computed by the OSFT in the trivial background $\mu = 0$
({\it i.e.}, with the usual kinetic term), but with the
addition  on an extra linear term $\mu X$.  

\smallskip

This is precisely what we would expect if the effect of deforming
the closed string background to $\mu \neq 0$ was captured by an open/closed vertex linear in the open string field.
In fact, it is well-known  that OSFT can reproduce amplitudes with
closed string insertions and at least a boundary by adding to the
action an appropriate open-closed vertex \cite{openclosed}, a linear term coupling
the closed string vertex operators to the open string fields. 
Since the cosmological constant operator is $Q_S$-closed, the open-closed vertex does not ruin the
topological localization, and reduces exactly to $ \frac{\mu}{g_s} \Tr X$
in the matrix integral!  In more complicated string theories, we would not expect in general
to be able to exponentiate
 a finite defomation of the closed string background
 by simply adding this linear term, but evidently
 this procedure is justified here. In particular
 cosmological constant operator ${\cal O}_1$ does not have contact terms with itself
 which would obstruct a naive exponentiation.

\smallskip

We see here what may well be the simplest illustration
of background independence in string field theory. We can either
start from the trivial background $\mu = 0$ and shift $\mu$ 
through the open/closed vertex, as in \refb{ZX}, or formulate
directly the theory in the new background with $\mu \neq 0$,
as in  \refb{Znew}. Background indipendence dictates that
the two forms of the action must be related
by a field redefinition, which in this case
is just a linear shift of  the ``string field'' $X$.

\smallskip

Is it possible to turn on other sources $t_k$ using the same procedure? For $t_3$,
corresponding to the dilaton operator ${\cal O}_3$, 
the  Ward identity can be integrated in a similar way and it simply
gives an appropriate rescaling of the relation between $\mu_B$ and
$z$. This is equivalently expressed
by adding to the matrix action the simple open-closed vertex $
\frac{3 t_3}{g_s} \Tr Z^2 X$, just as expected.
This procedure is not expected to work as easily
for higher $t_k$'s, as the operators
now have a non-trivial algebra of contact terms. Rather one
may anticipate a complicated matrix action containing
multi-trace interactions.

\smallskip
Finally we should briefly outline how the analysis of this
section could be recast in the language of integrable hierarchies.
Turning on $\mu$ corresponds to moving in the ``small phase space''
(which for the (2,1) model contains only the operator ${\cal O}_1$). The
relation between the KP times $t_k$ and the coordinates
$\{ z_i \}$ changes according to well-known formulas (see {\it e.g.} sections 4.2-4.3 of \cite{Dreview})
which could have been used to deduce the relation \refb{mod}.
Here we have phrased the discussion in a perhaps more 
intuitive physical language.

\sectiono{Future directions}

There are many  interesting directions in which the work of this
paper may be continued. In this section
we mention some of them.

\subsection{Relation with discretized random surface in ${\bf D=-2}$}

In this paper we have focused on
the Kontsevich model for the $(2,1)$ string theory.
There is also  a double-scaled matrix model for this closed string theory,
defined in terms of a matrix $M(\theta^1, \theta^2)$
that depends of two Grassmann-odd coordinates
\cite{minus2, KW1, KW2, KWloops, plefka}.
This model has a rich structure with many intriguing properties.

\smallskip

In the continuum limit, the coordinates $\theta^1$ and $\theta^2$
become precisely our fields $\Theta^\alpha$. This is one
of the reasons why one should prefer
the $\Theta^\alpha$ system to the $\xi \eta$ system.
Following the philosophy of \cite{JH},
this doubled-scaled matrix model should be understood
as the open string field theory on {\it unstable} D-branes
of the theory. Indeed, if one considers in the continuum
$(2,1)$ string theory  ZZ boundary conditions for the Liouville direction, and Neumann
b.c.  for the $\Theta^\alpha$ system, one finds
that the tachyon dynamics is captured
by a matrix $M(\theta^1_0,  \theta^2_0)$,
where $\theta^\alpha_0$ are the zero-modes
of $\Theta^\alpha$ living on the Neumann boundary.

\smallskip

\smallskip
In \cite{KWloops},  macroscopic loop operators for
this matrix model are considered. The operators of topological
gravity appear to be related to loop operators with Dirichlet boundary conditions on  the $\theta^\alpha$.
This seems to agree with our construction, and it
would be nice to understand this connection in detail.

\smallskip

More generally, it is of interest to see whether
our approach can shed some light on open/closed duality \cite{JH}
for the double-scaled matrix models. In the
``old'' approach, the doubled-scaled matrix model
is thought of as a trick to discretize the Riemann surface,
and it is essential to send $N$ to infinity and $t \to t_c$
to recover the continuum theory.  The modern approach
starts instead from considering the worldvolume
theory of a finite number $N$ of ZZ branes
in the continuum string theory. The precise relation
between the old and the new approach is still
quite unclear, as one cannot directly
identify the finite $N$ matrix model before
double-scaling limit with the finite $N$
open string field theory of the ZZ branes.
The OSFT of $N$ ZZ branes, with $N$ finite,
is presumably a  unique and consistent continuum quantum theory,
while the finite $N$ matrix model has
non-universal features, like the precise form of the potential.
The OSFT on $N$ ZZ branes may be expected \cite{Sen} to be dual to a subsector of the
full continuum closed string theory. This is
in analogy with the finite $N$ Kontsevich model.\footnote{We thank
Ashoke Sen for pointing out this analogy.}

\subsection{Generalizations }

The most obvious generalization of this work
that comes to mind is to the other $(p,q)$
minimal string theories. $(p, q)$ theories are solved
by double-scaling of the $(p-1)$-matrix chain,
where again $q$ labels the order of criticality.
$(p,1)$ models represent the ``topological points",
from which the  $(p,q)$ models with $q>1$
are obtained by flows of the $p$-KdV hierarchy.
There is a Kontsevich model  for any $(p,1)$ theory,
it is a one-matrix integral with a potential of
order $p+1$.  Our logic leads us to believe that
the OSFT on the stable branes of the $(p,1)$ theory
will localize topologically to a matrix integral.
Since OSFT is cubic, this process will lead to
a cubic matrix integral involving several matrices
(a matrix for each open topological primary).  The
simplest guess is that such cubic models are
related to the known   polynomial
Kontsevich models by integrating out
all matrices but one. A formulation in terms of a cubic multi-matrix
integral may have the advantage of making more
transparent the relation with a decomposition of
moduli space, which has not been completly understood for the
intersection numbers associated to the $(p,1)$ models.
Work is in progress along these lines.

\smallskip

Several other generalizations can be contemplated.
$\hat c < 1$ theories admit topological points and
to the best of our knowledge there is no known
topological matrix model description; our procedure should  give one.
The case of $c=1$ at the self-dual radius should also be attacked.

\sectiono{Conclusions}

In this paper we have described an example of  exact
open/closed duality that should represent the simplest
paradigm for  a large class of similar dualities. The
worldsheet picture of holes shrinking to punctures
is not, we believe, an artifact  of the simplicity of
the model, and the same mechanism may be at work in more physical situations.
We have found that at least in this example, open string field theory
on an infinite number of branes is capable of
describing the full string theory. This may contain
 a more general lesson.\footnote{Open string field theory on an infinite number
of branes has been  conjectured \cite{wittenK} to be relevant
for the issue of background independence in string theory.}
Although here we have stressed the importance of open
string field theory as a tool to understand open/closed
duality, one of our original motivations
was to learn about the structure of OSFT itself
in the solvable context of low-dimensional string theories.
The Kontsevich model is arguably the simplest imaginable
OSFT. It is still a good question
whether this and related examples can be used
to sharpen our understanding of OSFT.

\smallskip

We would like to conclude with a
speculation about how open/closed duality may come about in  AdS/CFT.
The example of the Kontsevich model suggests
that the natural starting point is the closed string theory
dual to free SYM ('t Hooft parameter $t = 0$).
At the point $t=0$, which
in some sense must
correspond to an infinitely curved AdS space,
 the closed string theory is expected to have an infinite
dimensional symmetry
group.
 This is analogous to the statement that $\{ t_k = 0 \}$ is the
topological point of the Kontsevich model.
If a a concrete description of this
closed string theory were available, one may also hope to define D-branes.
 D-branes of a peculiar nature may exist, such that:
1)The open string field theory on these D-branes is precisely the SYM theory,
with no extra massive open string modes. 2)When considered
in the closed string channel, the presence of the D-brane
can be completely re-adsorbed in a shift
of the closed string background. Adding D-branes
would then be equivalent  to turning on a finite
$t$, that is, to recovering a  smooth AdS space.
Statement 1) is analogous
to the topological localization that we have described
for the Kontsevich model, while statement 2) is the by now familiar
mantra of replacing boundaries with punctures.
This scenario would offer a derivation of AdS/CFT
 orthogonal to the usual one \cite{juan} that begins with D-branes in flat space and
proceeds by  ``dropping the one''  in the harmonic function.


\section*{Acknowledgments}

It is a pleasure to thank  Sunny Itzhaki,  Igor Klebanov,  John McGreevy,
Nathan Seiberg,  Joerg Teschner and  especially Ashoke Sen
and Herman Verlinde for very useful discussions and comments.
This material is based upon work supported by the National Science Foundation
Grant No. PHY-0243680.
Any opinions, findings, and conclusions or recommendations
expressed in this material are those of the authors and do not
necessarily reflect the views of the National Science Foundation.

\sectiono{Appendix: Liouville BCFT correlators}

In this appendix we give the
technical details of the computation
of 2- and 3-point vertices of open string tachyons.

\smallskip

We need the explicit expressions
of 2- and 3-point functions of  boundary primary operators
in Liouville BCFT (with FZZT boundary conditions).
The relevant formulas  can be found in \cite{FZZ, Kostov}.
We use the notations of \cite{FZZ}.
The variable $s$ is conventionally introduced\footnote{The FZZT
BCFT shows an interesting monodromy
in the complex $\mu_B$ plane \cite{teschnerMon}. The physics
is instead entire-analytic in terms of $s$.}
\be
 \frac{\mu_B^{}}{\sqrt{\mu}} =
\cosh{ b \pi s} \,.
\ee
Here $\mu$ is the bulk cosmological constant. We
are interested in the limit $\mu \to 0$, since this
is the point $\{ t_k = 0 \}$. Interestingly, the results for 2- and
3-point  correlators of open string tachyon turn out to be independent of $\mu$.

\smallskip

An important ingredient is the special function
${\bf G}_b(x)$ defined in \cite{FZZ}. This function
is entire-analytic and has zeros for $x=-n b -m/b$, with $m,n =0,1,2,\cdots$;
it is symmetric under $b \leftrightarrow 1/b$.
A convenient combination of ${\bf G_b}$'s is the
function ${\bf S}_b(x) = {\bf G}_b(Q-x) /{\bf G}_b(x)$, which obeys
the shift relation
\be
{\bf S}_b(x+b) = 2 \sin(\pi b x) {\bf S}_b(x) \,.
\ee
The 2-point function of boundary primary fields is then \cite{FZZ}
\ben \label{pre}
&& d({\alpha}, \mu_{B}^{(1)},\mu_{B}^{(2)},\mu) \equiv
\langle e^{\alpha \phi}  e^{\alpha \phi} \rangle =
(\frac{\pi}{\sqrt{2}} \mu
\gamma(\frac{1}{2}))^{\frac{3}{2}-{\sqrt{2} \alpha}} \times \\
&&  \quad \times \frac{{\bf G}_{\frac{1}{\sqrt{2}}}(-2{\alpha}+\frac{3}{\sqrt{2}})
{\bf S}_{\frac{1}{\sqrt{2}}}(\frac{3}{\sqrt{2}}+i(\,s_1
+\,s_2)/2-{\alpha}\,){\bf S}_{\frac{1}{\sqrt{2}}}(\frac{3}{\sqrt{2}}+i(\,s_1
-\,s_2)/2-{\alpha})}{{\bf G}_{\frac{1}{\sqrt{2}}}(-\frac{3}{\sqrt{2}}+2{\alpha})
{\bf S}_{\frac{1}{\sqrt{2}}}(i(\,s_1
+\,s_2)/2+{\alpha}){\bf S}_{\frac{1}{\sqrt{2}}}(i(\,s_1 -\,s_2)/2+{\alpha})} \,. \nonumber
\een
We now take $\alpha = b + \epsilon$. As $\epsilon \to 0$
 there is a pole arising from the zero of the first ${\bf G}_b$ in the
denominator.  The interesting residue is contained in the part of the
expression, finite for $\alpha \to b = \frac{1}{\sqrt{2}}$, that
contains the four ${\bf S}_{\frac{1}{\sqrt{2}}}$ functions,
 \ben
&& \frac{{\bf S}_{\frac{1}{\sqrt{2}}}(\frac{2}{\sqrt{2}}+i(\,s_1
+\,s_2)/2){\bf S}_{\frac{1}{\sqrt{2}}}(\frac{2}{\sqrt{2}}+(\,s_1 -\,s_2)/2)
}{{\bf S}_{\frac{1}{\sqrt{2}}}(i(\,s_1
+\,s_2)/2+\frac{1}{\sqrt{2}}){\bf S}_{\frac{1}{\sqrt{2}}}(i(\,s_1
-\,s_2)/2+\frac{1}{\sqrt{2}})} = \\
&& 4\sin(\frac{\pi}{2} + \frac{i \pi}{2\sqrt{2}}(s_1+s_2))
\sin(\frac{\pi}{2} + \frac{i \pi}{2\sqrt{2}}(-s_1+s_2)) =\nonumber \\
 && 2
\cosh(\frac{\pi}{\sqrt{2}} s_1) + 2 \cosh(\frac{\pi}{\sqrt{2}} s_2) = \nonumber
 2\, \frac{\mu_B^{(1)}+ \mu_B^{(2)}}{\sqrt{\mu}} \,.
\een
The factor of $1/\sqrt{\mu}$ cancels against the $\sqrt{\mu}$ in the prefactor of \refb{pre}.
This proves the claim \refb{pole}.

\smallskip

The three point function simplifies
when one takes the three Liouville momenta to be
equal to $b$. For generic $b$, this 3-point function
is proportional to a rational function of $\mu$, $\mu_B$ and
the ``dual''
cosmological constant $\tilde \mu^{}_B$ \cite{Kostov},
\be
\langle e^{b \phi }\, e^{b \phi} \, e^{b \phi} \rangle \sim
\frac{\tilde \mu_B^{(1)} ( \mu_B^{(2)}-\mu_B^{(3)} ) + \tilde
\mu_B^{(2)} ( \mu_B^{(3)}-\mu_B^{(1)} ) + \tilde \mu_B^{(3)} (
\mu_B^{(1)}-\mu_B^{(2)} ) }{( \mu_B^{(2)}-\mu_B^{(3)} )(
\mu_B^{(3)}-\mu_B^{(1)} )( \mu_B^{(1)}-\mu_B^{(2)} )}.
 \ee
For $c_{Liou} =28$, the dual cosmological constant obeys
\be \tilde \mu_B^{(i)} \sim  (2 (\mu_B^{(i)})^2 - \mu)
 \ee
and the tachyon 3-point function is just
a constant independent of $\mu$ and $\mu_B^i$.

\smallskip

Here we have computed the Liouville correlators
using analytic continuation in the Liouville momentum.
(Equally well, we could have use analytic continuation
in $b$ to regulate the expressions that become
singular as $b \to 1/\sqrt{2}$. Indeed one of the achievements
of the past few years has been the recognition that Liouville
correlators have nice analytic properties with respect
to all the parameters.) If one insists in working
strictly at $b=1/\sqrt{2}$ and with the on-shell
vertex operators $e^{b \phi}$, an alternative
way to phrase the results is the language of logarithmic CFT \cite{gurarie}.
For generic $b$, the two operators $e^{\alpha \phi}$ and
$e^{ (Q - \alpha) \phi}$ are identified as
\be
e^{\alpha \phi} = d({\alpha}, \mu_{B}^{(1)},\mu_{B}^{(2)},\mu)  \, e^{ (Q - \alpha) \phi} \,.
\ee
The reflection coefficient  $ d({\alpha}, \mu_{B}^{(1)},\mu_{B}^{(2)},\mu)$
has poles for $ Q- 2 \alpha =n b + m/b$.  For these cases,
the identification becomes ill-defined. One way around this
is that for these resonant values $\tilde \alpha$ we
 modify the identification as
\be
(L_0 - h_{\tilde \alpha}) \, e^{\tilde \alpha \phi} =  \left[
\lim_{\alpha \to \tilde \alpha }  (( h_{\alpha} - h_{\tilde \alpha})  d({\alpha}, \mu_{B}^{(1)},\mu_{B}^{(2)},\mu) ) \,
\right] e^{ (Q - {\tilde \alpha}) \phi}   \, .
\ee
Notice that the term is in square brackets is just a finite
coefficient. $L_0$ cannot be diagonalized in
the subspace spanned by  $e^{\tilde \alpha \phi}$ and  $e^{ (Q - {\tilde \alpha}) \phi} $,
which forms a non-trivial Jordan cell. In other terms,  the two operators
are a logarithmic pair. In our case, $\tilde \alpha = b$.
Working at $b$ strictly equal to $1/\sqrt{2}$, we can write
\be
L_0 \,  e^{\phi(0)/\sqrt{2}} c_1 | 0 \rangle_{ij} = (\mu_B^{(i)}  + \mu_{B}^{(j)} ) \,  e^{\sqrt{2} \phi(0)}  c_1 | 0 \rangle_{ij} \,.
\ee
This gives an alternative way to understand why the
tachyon kinetic term in the OSFT action is $ (\mu_B^{(i)}  + \mu_{B}^{(j)} )$.

\begingroup\raggedright


\endgroup

\end{document}